\documentclass[prd,floatfix,twocolumn]{revtex4}
\usepackage{hyperref}
\usepackage{natbib,ifthen}

\usepackage{amssymb} 
\usepackage{amsmath}
\usepackage{graphicx}
\usepackage{subfigure}

\usepackage{color}
\def\be{\begin{equation}}
\def\ee{\end{equation}}
\def\bea{\begin{eqnarray}}
\def\eea{\end{eqnarray}}

\def\lsim{\mathrel{\mathstrut\smash{\ooalign{\raise2.5pt\hbox{$<$}\cr\lower2.5pt\hbox{$\sim$}}}}}
\def\gsim{\mathrel{\mathstrut\smash{\ooalign{\raise2.5pt\hbox{$>$}\cr\lower2.5pt\hbox{$\sim$}}}}}

\def\({\left(}
\def\){\right)}

\def\bwt{\begin{widetext}}
\def\ewt{\end{widetext}}
\begin{document}
\bibliographystyle{prsty}
\title{N-Body Simulations of DGP and Degravitation Theories}
\author{Justin Khoury$^{1,2}$}
\author{Mark Wyman$^{2}$}
 \affiliation{$^{1}$ Department of Physics and Astronomy, University of Pennsylvania, Philadelphia, PA 19104, USA\\
$^{2}$ Perimeter Institute for Theoretical Physics, Waterloo, Ontario N2L 2Y5, Canada}
\date{\today}
%\preprint{hep-th/yymmnnn}
\begin{abstract}
We perform N-body simulations of theories with infinite-volume extra dimensions, such as the Dvali-Gabadadze-Porrati (DGP) model
and its higher-dimensional generalizations, where 4D gravity is mediated by massive gravitons. The longitudinal mode of these gravitons
mediates an extra scalar force, which we model as a density-dependent modification to the Poisson equation. This enhances gravitational
clustering, particularly on scales that have undergone mild nonlinear processing. While the standard non-linear fitting algorithm
of Smith {\it et al.} overestimates this power enhancement on non-linear scales, 
we present a modified fitting formula that offers a remarkably
good fit to our power spectra. Due to the uncertainty in galaxy bias, our results are consistent with precision
power spectrum determinations from galaxy redshift surveys, even for graviton Compton wavelengths as small as 300~Mpc. 
Our model is sufficiently general that we expect it to capture the phenomenology of a wide class of related
higher-dimensional gravity scenarios. 
\end{abstract}
\maketitle
%%%%%%%%%%%%%%%%%%%%%%%%%%%%%%%%%%%%%%%%
\section{Introduction}
%%%%%%%%%%%%%%%%%%%%%%%%%%%%%%%%%%%%%%%%

One of the most pressing questions in cosmology is whether the mounting evidence for dark energy
is in fact a consequence of a breakdown of Einstein's gravity on the largest scales. While observations
are converging on a background history consistent with that predicted by $\Lambda$-Cold Dark Matter
($\Lambda$CDM) cosmology, the most stringent tests on the standard gravity/$\Lambda$CDM
paradigm will come from the formation of large scale structure~\cite{bhuvnesh}.

In this paper we study structure formation within a well-motivated class of infrared-modified gravity
theories, namely those in which 4D gravity is mediated by massive gravitons with characteristic mass $r_c^{-1}$. Aside from
being phenomenologically interesting to study, these theories shed new light on the cosmological
constant problem~\cite{weinberg}. Because the graviton is massive, long wavelength sources
--- such as vacuum energy --- may effectively decouple from gravity, or degravitate~\cite{dilute,nimagia,degrav}.

The simplest and best-studied example in this class is the Dvali-Gabadadze-Porrati (DGP) scenario~\cite{DGP}, consisting of a 3-brane with one extra dimension. Since the bulk is flat and infinite in extent, gravity on the brane does not reduce to general relativity at low energies. Instead, the force law is approximately $1/r^2$ at short distances but weakens to $1/r^3$ at distances much greater than $r_c$. Much effort has been devoted recently to confronting DGP with
cosmological observations~\cite{lue,mustafa,koyama,hu,hu2,wiley,amin,song,columbia,Laszlo:2007td}, most of which pertains to the self-accelerated or unstable branch.
In this work we focus exclusively on the {\it normal} or {\it stable} branch of DGP.

\begin{figure}[h!] %  figure placement: here, top, bottom, or page
   \centering
   \includegraphics[width=0.5\textwidth]{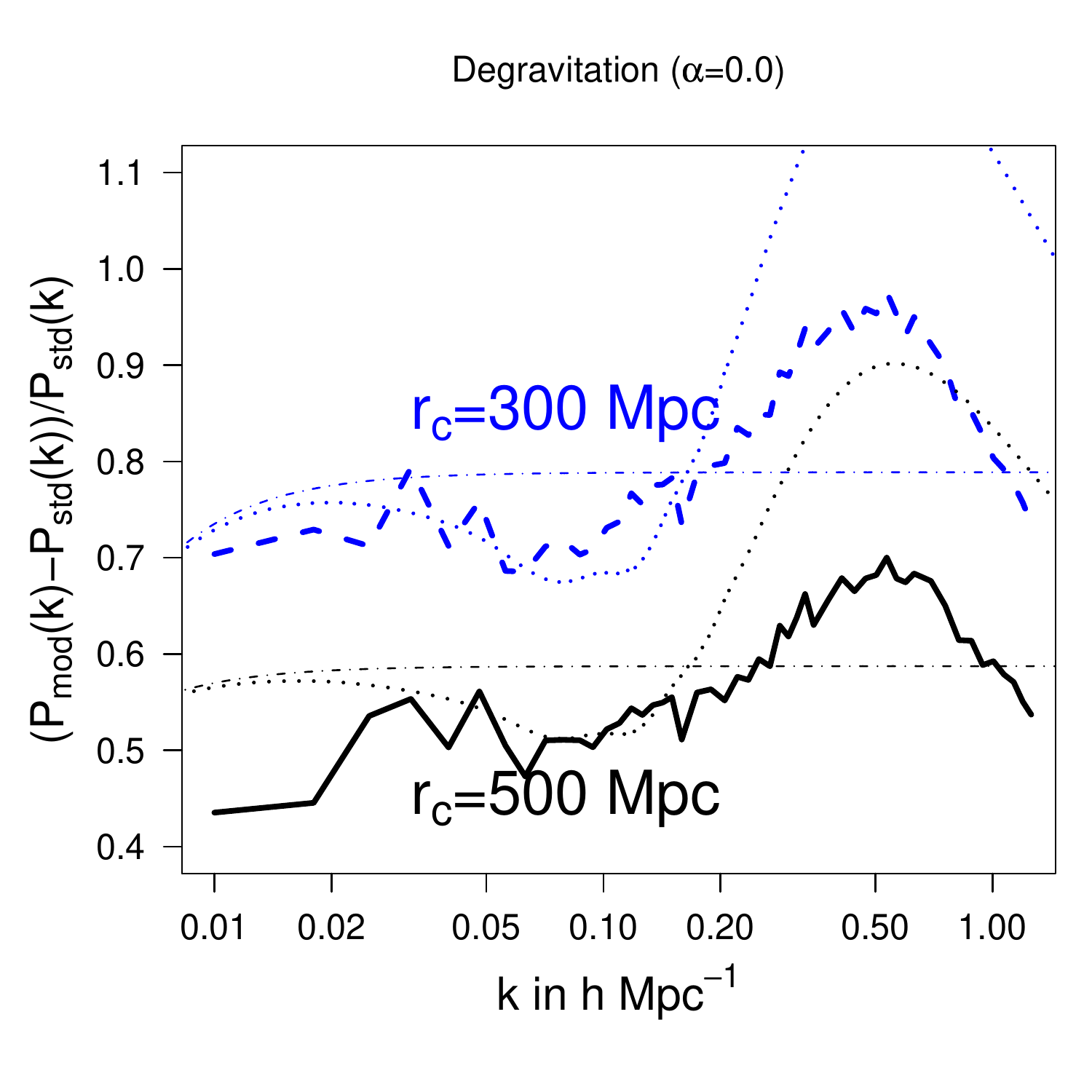} 
   \caption{Fractional difference between the power spectra for degravitation/cascading models ($\alpha = 0$) with
   graviton Compton wavelength of $r_c=300$ Mpc (blue dashed) and $r_c = 500$ Mpc (black solid) and that of standard gravity, without normalizing to data. The dash-dotted line is the expected difference from linear perturbation theory. The dotted line is the expected difference assuming the Smith {\it et al.} procedure~\cite{Smith:2002dz} for including nonlinear effects. }
   \label{fracdiff}
\end{figure}

Our analysis more generally encompasses extensions of DGP to higher dimensions --- 
a class of models that are non-self accelerating and free of ghost-like instabilities. In particular, 
we are interested in the cascading gravity framework~\cite{oriol,us,claudiareview}. In this construction, our 3-brane lies within a succession of higher-dimensional DGP branes, embedded in one another within a flat bulk space-time~\cite{oriol,us}. This hierarchy of branes in turn leads to a force law that successively goes through a $1/r^2$ regime, followed by $1/r^3$, then $1/r^4$ etc., as one probes larger distances from a source. (A similar cascading behavior was also obtained through a different construction in~\cite{nemanjacharting}.) It was recently argued~\cite{niayeshghazal} that the cosmological predictions of cascading gravity models can explain certain anomalies in the data. 

When massive, gravitons propagate 5 polarization states: the 2 helicity-2 states of Einstein gravity,
2 helicity-1 states, and 1 helicity-0 or longitudinal mode. The last, traditionally denoted by $\pi$, is most interesting phenomenologically ---
it mediates an extra scalar force which enhances structure growth. However, this extra force is only effective
at sufficiently low density: thanks to the Vainshtein effect~\cite{vainshtein}, non-linearities decouple $\pi$ near astrophysical sources, thereby insuring consistency with solar system constraints. (This is qualitatively similar to the chameleon mechanism \cite{cham1,cham2,cham3}, at play in phenomenologically consistent examples~\cite{f(R)wayne} of $f(R)$ gravity models~\cite{f(R)}.) Translated to the cosmological context, this screening mechanism implies
that gravity becomes stronger only at late times and on sufficiently large scales.

\bwt

\begin{figure}[htbp] %  figure placement: here, top, bottom, or page
   \centering
   \includegraphics[width=\textwidth]{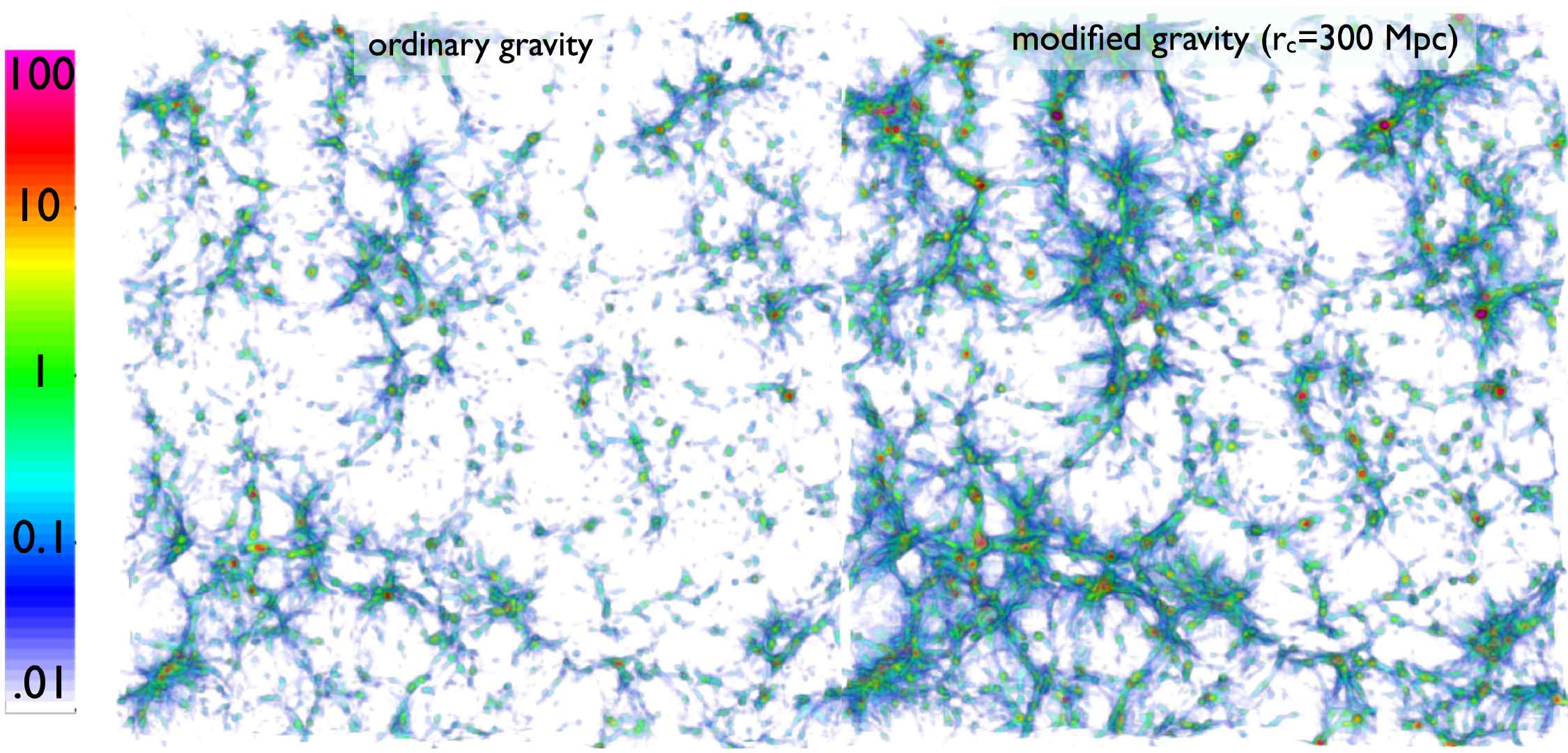} 
   \caption{Kinetic energy density at $z=0$ in a slice of depth 31.25~$h^{-1}$Mpc from a pair of 400~$h^{-1}$Mpc simulations with the same initial conditions, evolved according to standard gravity on the left and a degravitation/cascading model ($\alpha = 0$, $r_c = 300$ Mpc) on the right. The units displayed on the scale are arbitrary but common to both panels. We plot kinetic energy density, rather than simple overdensity, because the 
 density enhancement in these models is too subtle to detect readily by eye in such a plot. Kinetic energy density, however, is enhanced significantly by the greater gravitational force felt by the particles.}
   \label{keden}
\end{figure}

\ewt

\begin{figure}[h!] %  figure placement: here, top, bottom, or page
   \centering
   \includegraphics[width=0.5\textwidth]{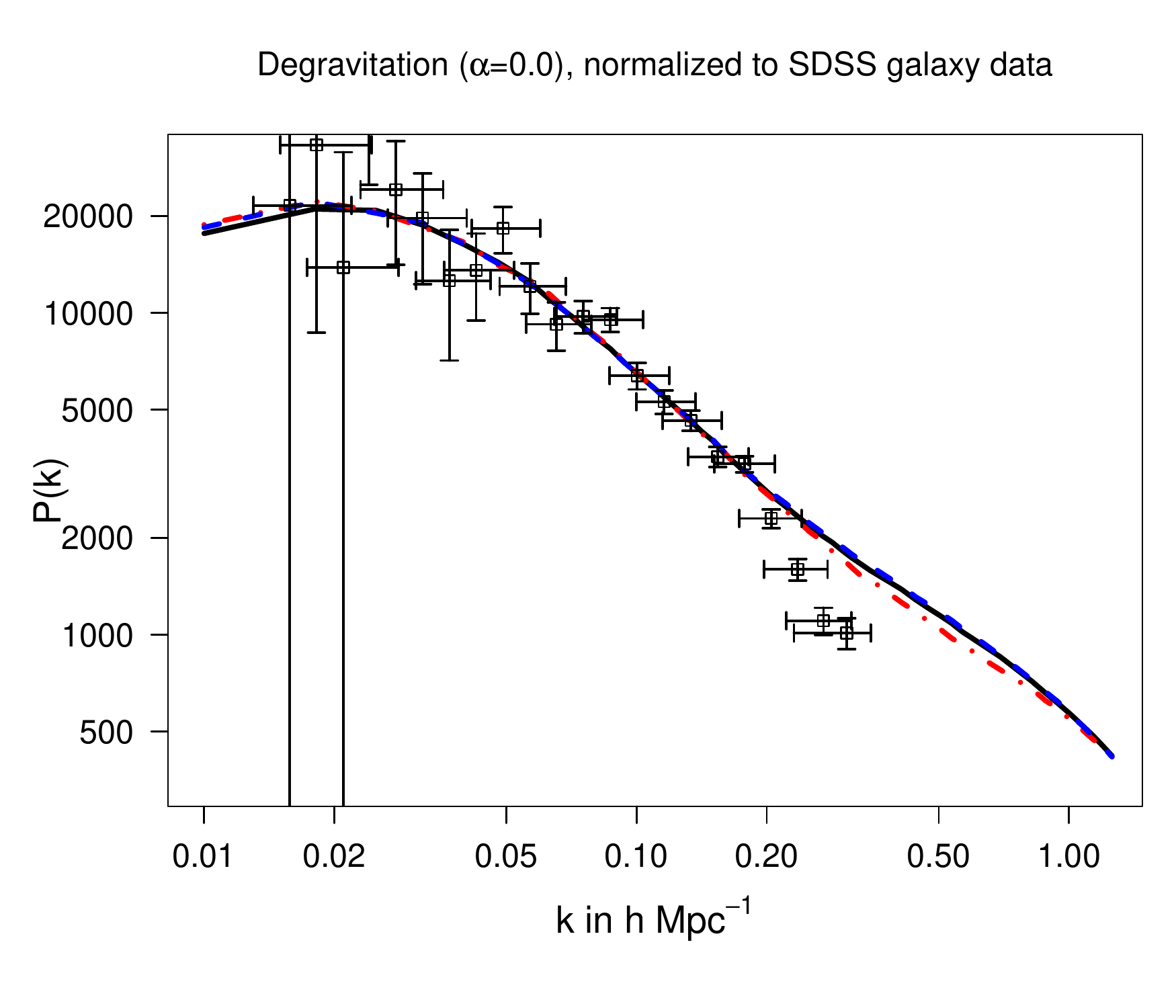} 
   \caption{Power spectra for $r_c=300$ Mpc (blue dashed), $r_c = 500$ Mpc (black solid) and standard gravity (red dash-dotted), each separately normalized to the Sloan Digital Sky Survey main galaxy data~\cite{sloan} (data points).This figure demonstrates that the apparent difference among the power spectra, as viewed via galaxy redshift surveys, is small, given the uncertainty in galaxy bias. Compared with the bias of the standard gravity power spectrum, the modified spectra biases for $r_c=$ 300 (500)~Mpc are 57\% (65\%) relative to that needed for the standard gravity results.}
   \label{vdata}
\end{figure}

\subsection{Summary of Results}

To perform N-body simulations of DGP and cascading/degravitation models, we work in the Newtonian limit where
the perturbation equations are local on the brane. In the DGP context, these have been worked out explicitly
by Ref.~\cite{lue} for the case of a spherical top-hat perturbation. From the parametric dependence of these results, we can infer their generalization
to higher-dimensional cascading models. For simplicity, we assume a background expansion history identical to
that of $\Lambda$CDM cosmology, allowing us to focus on the effects from the modified growth history. Moreover, the $\Lambda$CDM background
cosmology is a realistic approximation for the expansion history expected in higher-dimensional DGP models~\cite{niayeshghazal}, as we review below.

Figure~\ref{fracdiff} shows the relative difference in power spectrum between cascading/degravitation models and standard gravity, with $r_c = 300$ (dashed blue line) and $500$~Mpc (black solid line). The spectra from our simulations, spanning the range $0.01 \lesssim k  \lesssim1.2 \; h\, {\rm Mpc}^{-1}$, have been extended to smaller $k$ using results from direct numerical solution of the linear perturbation equations. Because of the extra scalar force mediated by $\pi$, 
gravitational clustering is enhanced substantially on a wide range of scales, and further enhanced on
scales that have undergone nonlinear processing. And because this extra force becomes active at earlier times for smaller $r_c$,
the enhancement is greater for $r_c=300$ than $500$~Mpc. For comparison, we have also plotted the enhancement expected from linear theory (dash-dotted lines)
and the Smith {\it et al.} algorithm~\cite{Smith:2002dz} to account for non-linear effects (dotted lines). 

The enhancement of large scale structure growth in our modified gravity simulations is
seen directly by making a contour plot of the kinetic energy density in degravitated and standard gravity 
simulations. Figure~\ref{keden} compares the results at $z=0$ for slices of depth 31.25~$h^{-1}$Mpc
from 400 $h^{-1}$ Mpc simulations. Structure is clearly more evolved in the modified gravity panel
due to the $\pi$-mediated force. We plot kinetic energy density, rather than simple overdensity, because the 
 density enhancement in these models is too subtle to detect readily by eye in such a plot. Kinetic energy density, however, is greatly enhanced by the greater gravitational force felt by the particles in the simulation.

Due to uncertainties in the bias between galaxy redshift surveys and the CDM power spectrum we model, however,
this enhancement is nevertheless consistent with power spectrum determinations such as the Sloan 
Digital Sky Survey (SDSS). Figure~\ref{vdata} shows the fit to the Sloan main galaxy power spectrum~\cite{sloan}, assuming a scale-independent bias. While the fiducial cosmological parameters assumed here ---  $\Omega_{\rm m} = 0.3$, $n_s=1.0$, $h=0.7$ --- offer a poor fit to the data, even for the standard $\Lambda$CDM model, the purpose of this comparison is to demonstrate that the apparent difference in $P(k)$, as viewed via galaxy redshift surveys, is small, given the uncertainty in galaxy bias. For the record, the bias required for our $r_c = 300\; (500)$~Mpc models to fit these data is 56\% (67\%) of the bias parameter needed for the standard gravity results from our simulations. This is a consequence of the greater amplitude of the modified gravity CDM power spectra. Note that an important assumption in this comparison is that of a constant bias. It was argued recently that modified gravity theories generically lead to a scale-dependent bias~\cite{lam}, an effect we are currently quantifying using our simulations.

\section{Phenomenology of Massive/Resonance Gravity}
\label{DGPreview}

\subsection{Fundamentals}

The modified gravitational law in standard DGP follows from the graviton propagator $1/(k^2 + r_c^{-1}k)$, where $r_c$ sets the cross-over scale
from the 4D ($1/r^2$) to the 5D ($1/r^3$) regime. While the more complicated nature of the force law in higher-dimensional degravitation models does not lend itself to such a simple form for the propagator, a useful parameterization is~\cite{giaalpha,degrav}
\be
\frac{1}{k^2 + r_c^{-2(1-\alpha)}k^{2\alpha}}\,,
\label{prop}
\ee
with standard DGP corresponding to $\alpha = 1/2$. This power-law parameterization is not only simple in form, but it also makes contact with
the far infrared limit of the cascading/degravitation propagator: since these models all have $D\geq 6$ space-time dimensions, the force law scales as $1/r^{D-2}$ in the far infrared, corresponding to a propagator that tends to a constant ($D>6$) or behaves as $\log k$ ($D=6$) as $k\rightarrow 0$. Thus all higher-dimensional extensions of DGP correspond to $\alpha\approx 0$ theories in the infrared~\cite{us}.

In this work, therefore, we shall be primarily interested in $\alpha=1/2$ and $\alpha=0$, corresponding respectively to standard DGP and cascading/degravitation models.
More generally, $\alpha$ is a free parameter within the allowed range $0 \leq \alpha < 1$. The upper bound is required in order for the modification to be relevant in the infrared; the lower bound follows from unitarity~\cite{giaalpha}. 

The above propagator describes a resonance graviton --- a continuum of massive gravitons --- whose spectral density peaks at the tiny scale $r_c^{-1}$.
By virtue of being massive spin-2 particles, each graviton state propagates 5 polarizations, including a helicity-0 or longitudinal mode $\pi$.
(Cascading gravity models also have
extra 4D scalar degrees of freedom inherited from the higher-dimensional massless graviton~\cite{oriol,nonFP}. For simplicity, in this work
we shall ignore these scalars and focus exclusively on $\pi$.)

As emphasized above, this longitudinal mode is responsible for nearly all of the interesting phenomenology of the models considered here. At the linearized level, it contributes an
additional $T^\mu_{\; \mu}T'^\nu_{\;\nu}/6$ to the one-particle exchange amplitude between conserved sources, which at first sight
would seem grossly to violate solar system constraints~\cite{vDVZ}. 

As Vainshtein~\cite{vainshtein} realized, however, the weak-field approximation is invalid for the longitudinal mode in the vicinity of astrophysical sources. 
Instead, non-linearities in $\pi$ become important and result in it decoupling from matter on scales smaller
than a macroscopic scale given by~\cite{giaalpha}
\be
r_\star = \left(r_c^{4(1-\alpha)}r_{\rm Sch}\right)^{1/(1+4(1-\alpha))}\,,
\label{rstar}
\ee
where $r_{\rm Sch}$ is the Schwarzschild radius of the source. For $\alpha = 1/2$, this $r_\star$-effect has been confirmed in explicit
DGP solutions~\cite{ddgv,gruz,por}.

To be more explicit, on scales $r\ll r_\star$ the leading correction to the Newtonian potential is~\cite{giaalpha,degrav}
\be
\left. \frac{\delta\Psi}{\Psi}\right\vert_{r\ll r_\star} \sim \sqrt{\frac{r}{r_{\rm Sch}}}\left(\frac{r}{r_c}\right)^{2(1-\alpha)}\,.
\label{modpot}
\ee
The above parametric dependence is fixed by two requirements: $i)$ that $\pi$ be of order $r_{\rm Sch}/r_\star$ at $r=r_\star$, to match the linear solution;
$ii)$ that the solution be analytic in $r_c^{-2(1-\alpha)}$ for $r\ll r_\star$. Equation~(\ref{modpot}) shows that, as desired, $\pi$ leads to a small correction
to the Newtonian potential for $r\ll r_\star$. Thus non-linearities lead to a decoupling of $\pi$ and recovery of Einstein gravity locally.

For $r\gg r_\star$, on the other hand, $\pi$ yields a correction of order unity,
\be
\left.\frac{\delta\Psi}{\Psi}\right\vert_{r\gg r_\star}  = \frac{1}{3}\,.
\label{weak}
\ee
This is consistent with the $\pi$ contribution to the exchange amplitude mentioned earlier.

Thus one can think of the helicity-0 mode as mediating a scalar fifth force which is suppressed near high density sources
but becomes relevant at astrophysically large distances. This is precisely the opposite of a fifth force mediated
by a massive scalar field, where the Yukawa potential fades away at distances larger than
the Compton wavelength. The $\pi$ behavior instead closely resembles that of the $f(R)$ scalar field~\cite{f(R)} under the 
chameleon mechanism~\cite{cham1,cham2,cham3}. Indeed we will see that many of our results share qualitative features
of N-body simulations of $f(R)$ gravity~\cite{marcos}.

The key point is that $r_\star$, while large for astrophysical sources, is nevertheless parametrically smaller than $r_c$.
Quantitatively, a $10^{10}{\rm M}_\odot$ galaxy has a Vainshtein radius of $r_\star \sim 50\;{\rm kpc}$ for
$r_c = 300\;{\rm Mpc}$ with $\alpha = 0$. For a $10^{15}{\rm M}_\odot$ galaxy cluster, this gives $1\;{\rm Mpc}$. 
Given these scales, we expect $\pi$ to play an important role in structure formation.

\subsection{Background Cosmology}

The infrared modifications of gravity discussed above should translate in the cosmological context into corrections to the
expansion history. This is certainly the case in the standard DGP scenario, where the Friedmann equation
receives an added contribution proportional to $H/r_c$~\cite{cedric}. By virtue of being linear in $H$, however, this correction
results in too large a deviation from $\Lambda$CDM expansion history, leading to significant tensions
with current data~\cite{columbia}.

The situation is much more hopeful in higher dimensional models. As argued in~\cite{niayeshghazal}, for general $\alpha$ theories
we expect power-law corrections to the Friedmann equation of the form
\be
H^2 = \frac{8\pi G}{3}\rho + \frac{\Lambda}{3} - \frac{H^{2\alpha}}{r_c^{2(1-\alpha)}}\,.
\label{modfried}
\ee
(Recall that in this work we focus exclusively on the so-called ``normal" branch, as opposed to the self-accelerated branch,
hence the choice of minus sign on the right hand side. The ``plus" branch version of this equation was introduced in~\cite{turner}
to study generalized self-accelerated solutions.) In particular,~(\ref{modfried}) agrees with the standard DGP Friedmann equation~\cite{cedric} for $\alpha=1/2$.
More importantly, we immediately see that $\alpha=0$ (corresponding to 2 or more extra dimensions) leads to an expansion history identical to that of $\Lambda$CDM cosmology. Of course, a complete understanding of higher-codimension cascading gravity models will inevitably uncover small departures in their expansion history compared with $\Lambda$CDM, as they do not exactly correspond to $\alpha = 0$. But $\alpha = 0$ should offer a realistic approximation to their modified Friedmann equation since we expect the departures to be a slowly-varying function of $H/r_c$~\cite{claudiaandrew}.

In light of these considerations, in this work we therefore assume that {\it the expansion history is identical to that of $\Lambda$CDM cosmology.} 
This has the virtue of disentangling the effects due to a modified growth history, which is our primary interest.
And while this is only justified for higher-dimensional cascading models, we also assume a $\Lambda$CDM background expansion for DGP, 
so that our simulations can enlighten us on the $\alpha$-dependence of the modified growth history.

\subsection{Cosmological Perturbations}
\label{CP}

For the purpose of N-body simulations, we are interested in sub-Hubble scales and non-relativistic sources. In this regime, Lue {\it et al.}~\cite{lue} derived
the modified evolution equations for a spherical top-hat perturbation in the standard DGP model; these results
were extended in~\cite{Koyama:2007ih}. Here we propose a phenomenological extension of these results valid for general $\alpha$ theories, which reduce to the equations of~\cite{lue} in the DGP case.

The change in the perturbation equations can be entirely encoded as a scale-dependent modification to the Poisson equation:
\bwt
\be
\left(\frac{k^2}{a^2} + \frac{1}{r_c^2} \( \frac{k r_c}{a}\)^{2\alpha}\right)\Psi = -4\pi G \bar{\rho} \left \{ \left(1-\frac{2g}{\epsilon}\left[\sqrt{1+\epsilon}-1\right] \right)\delta\, \right \}_k,
\label{modpoisson}
\ee
\ewt
where $\epsilon$ measures the overdensity $\delta\equiv \delta\rho/\bar{\rho}$ in units of $r_c$,
\be
\epsilon = 8g^2r_c^2\frac{8\pi G}{3}\bar{\rho}\;\delta = 8g^2H^2r_c^2\Omega_{\rm m}\delta\,.
\label{eps}
\ee
and $g$ is defined by
\be
g = -\frac{1}{3}\cdot\frac{1}{1 + 2 \left(Hr_c\right)^{2(1-\alpha)}\left(1+ \frac{\dot{H}}{3H^2} \right)}\, ,
\label{g}
\ee
where the dot indicates a derivative with respect to proper time. Note that the right hand side of Eqn. \ref{modpoisson} is meant first
to be evaluated in real space before being transformed into Fourier space to match the form of the left hand side.

A few comments are in order: 

\begin{itemize}

\item The above results all agree with~\cite{lue} for the DGP case $\alpha =1/2$, with one exception: we have taken into account the modified propagator~(\ref{prop}) in the right-hand side of~(\ref{modpoisson}). This mass term has been neglected in most earlier studies of DGP cosmological perturbations~\cite{lue,koyama,song}, except in~\cite{amin}, presumably because $r_c$ was assumed to be of order $H_0^{-1}$. In this work, however, we shall consider much smaller values for $r_c$, and the mass term cannot be dropped.

\item The correction factor on the right hand side of~(\ref{modpoisson}) encodes the ``fifth force" mediated by the longitudinal mode $\pi$. This is most easily seen by considering the linearized approximation, $\epsilon\ll 1$, in which~(\ref{modpoisson}) simplifies to
\be
\left(\frac{k^2}{a^2} + \frac{1}{r_c^2}\left(\frac{kr_c}{a}\right)^{2\alpha}\right)\Psi =-4\pi G\bar{\rho} (1- g)\delta\,.
\label{lin}
\ee
The effective correction to Newton's constant, $G_{\rm eff} = G(1-g)$, can be attributed to the behavior of $\pi$ in a cosmological background. 
Indeed, for the universe as a whole both $r$ and $r_{\rm Sch}$ can be approximated by the Hubble radius, $H^{-1}$. Thus, in the strong coupling regime, $Hr_c\gg 1$, we can translate~(\ref{modpot}) to the cosmological context:
\be
\left.\frac{\delta\Psi}{\Psi}\right\vert_{Hr_c\gg 1} \sim \frac{1}{(Hr_c)^{2(1-\alpha)}}\,.
\label{picosmo2}
\ee
The parametric dependence agrees precisely with~(\ref{g}) in the $Hr_c\gg 1$ limit: $g(Hr_c\gg 1) \sim 1/(Hr_c)^{2(1-\alpha)}$. Similarly, in the weak coupling regime, we have $g(Hr_c\ll 1) = -1/3$, which is consistent with~(\ref{weak}).

\item The non-linear dependence on $\epsilon$ in~(\ref{modpoisson}) encodes the Vainshtein effect for local overdensities.
Indeed, for sufficiently large overdensities such that $\epsilon \gg 1$, the right-hand side of~(\ref{modpoisson}) reduces to its
standard gravity form. The dependence of the effective gravitational force as a function of $\delta$ is shown in Fig.~\ref{antiV} for $z=1$.
Because these are plotted at $z=1$, the enhancement at small $\delta$ is not yet maximized, thanks to the global Vainshtein effect.
In other words, $g$ has not yet relaxed to its late-time value of $-1/3$. Since $g$ is a slower function of $Hr_c$ for $\alpha = 1/2$ than for $\alpha = 0$,
the relaxation time is longer for $\alpha = 1/2$, hence the DGP curve has smaller amplitude at $z=1$.

Figure~\ref{antiV} underscores the fact that our modification depends on the local density, in a way akin to the chameleon mechanism in $f(R)$ gravity simulations~\cite{marcos}. This is unlike earlier studies that assumed density-independent modifications, such as~\cite{Stabenau:2006td,Dore:2007jh}.

\begin{figure}[htbp] %  figure placement: here, top, bottom, or page
   \centering
   \includegraphics[width=0.5\textwidth]{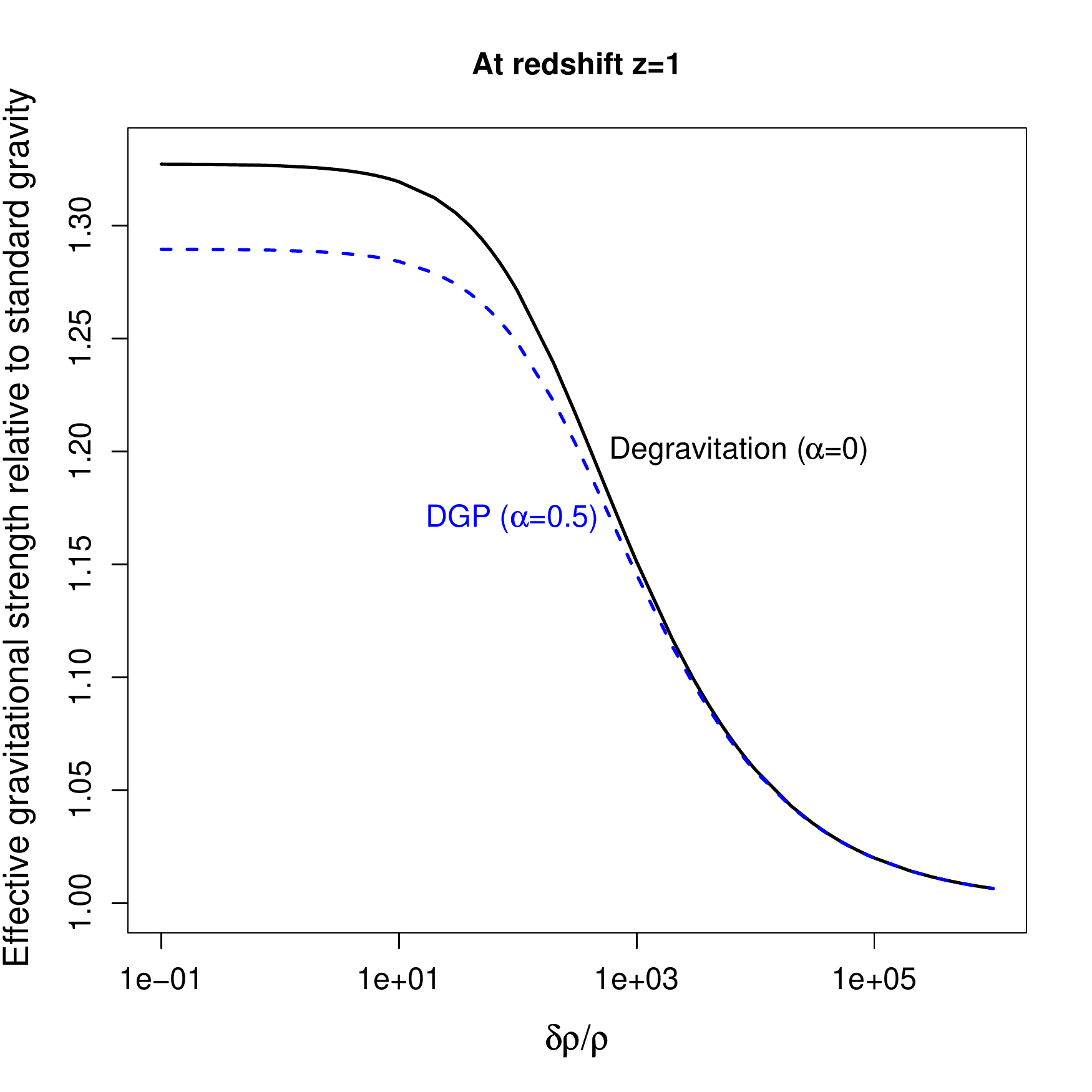} 
   \caption{Effective gravitational attraction relative to standard gravity at $z=1$ as a function of $\delta\rho/\rho$.  More precisely, what is plotted is $1-2g(\sqrt{1+\epsilon}-1)/\epsilon$, which appears on the right-hand side of~(\ref{modpoisson}). The enhancement compared to standard gravity fades away for large $\delta\rho/\rho$, a manisfestation of the Vainshtein screening effect.} 
\label{antiV}
\end{figure}

\item This same non-linear dependence also encodes a weak, but interesting, ``anti-Vainshtein" effect in local underdensities, where the gravitational force can become even a bit larger than even its usual low-overdensity enhanced value. In other words, if Fig.~\ref{antiV} were extended to include underdensities, the curves would keep rising slightly in the range $\delta < 0$. This effect merits further study, but could be relevant to the void phenomenon, as we will discuss briefly in Sec.~\ref{conc}.

\item The analysis of~\cite{lue}, on which the above results are based, assumes spherical top-hat perturbations. For general perturbations, the
equation of motion for $\pi$ in DGP takes a more complicated form and is therefore harder to solve. The precise relation between our framework and the $\pi$
equation of motion is spelled out in Appendix~\ref{decoup}. Ongoing N-body simulations in DGP
by R.~Scoccimarro have established that a spherical approximation yields a power spectrum that agrees well with the full calculation~\cite{roman}.

\item Implicit in our modified Poisson equation is some coarse-graining for the density field. Otherwise, since $\epsilon$ is formally a delta function for a point particle,
our modification would vanish everywhere in this extreme case. Of course, since~(\ref{modpoisson}) depends non-linearly on $\delta$, its solution will unavoidably have some sensitivity to the choice of coarse-graining scale. As we discuss further in Sec.~\ref{powspec}, however, we will check that this effect is under control by comparing simulations of different resolution. Note that these averaging issues also pertain to $f(R)$ gravity simulations~\cite{marcos}, since the chameleon mechanism depends sensitively on the size and density of objects.

\end{itemize}

Equations~(\ref{modpoisson})-(\ref{g}) constitute the core ingredients for our simulations. Given a density field $\delta$ and Hubble parameter $H$, we can solve~(\ref{modpoisson}) for the Newtonian potential. The latter then specifies how particles evolve in time in the usual way.

We should stress that the above expressions only hold well inside the horizon, which is of course all we care about for our simulations. In DGP, for instance, the corrections to general relativity take on a rather different form on super-Hubble scales, as shown by~\cite{hu,hu2} and~\cite{song} in the self-accelerated and normal branch, respectively. 

The generalization of~(\ref{lin}) for near- and super-Hubble modes, and the implications for large-scale growth history, has been studied recently in~\cite{niayeshghazal}. In particular, it was shown that modified gravity can explain the lack of CMB temperature correlation on $\gsim 60\,^{\circ}$ angular separations~\cite{wmap1,huterer,others}, provided that the curvature perturbation $\zeta$ is conserved on all relevant scales. The Newtonian limit
in this case is different than~(\ref{modpoisson}) , but the physics is qualitatively very similar --- enhanced structure growth on scales smaller
than the graviton Compton wavelength. 

We follow a more conservative approach here. By focusing on only two parameters, $r_c$ and $\alpha$, we are able to succinctly explore the large scale structure phenomenology of this class of models. Furthermore, since the modifications we introduce are less dependent on the somewhat uncertain super-horizon physics
of these extra-dimensional scenarios, we can expect our general conclusions to be robust to changes in those scenarios' particulars. As we learn more from observations and how they compare with simulations, we will be able to place better constraints on explicit brane constructions.

\subsection{Linear Regime}

\begin{figure}[h!] %  figure placement: here, top, bottom, or page
\includegraphics[width=\columnwidth]{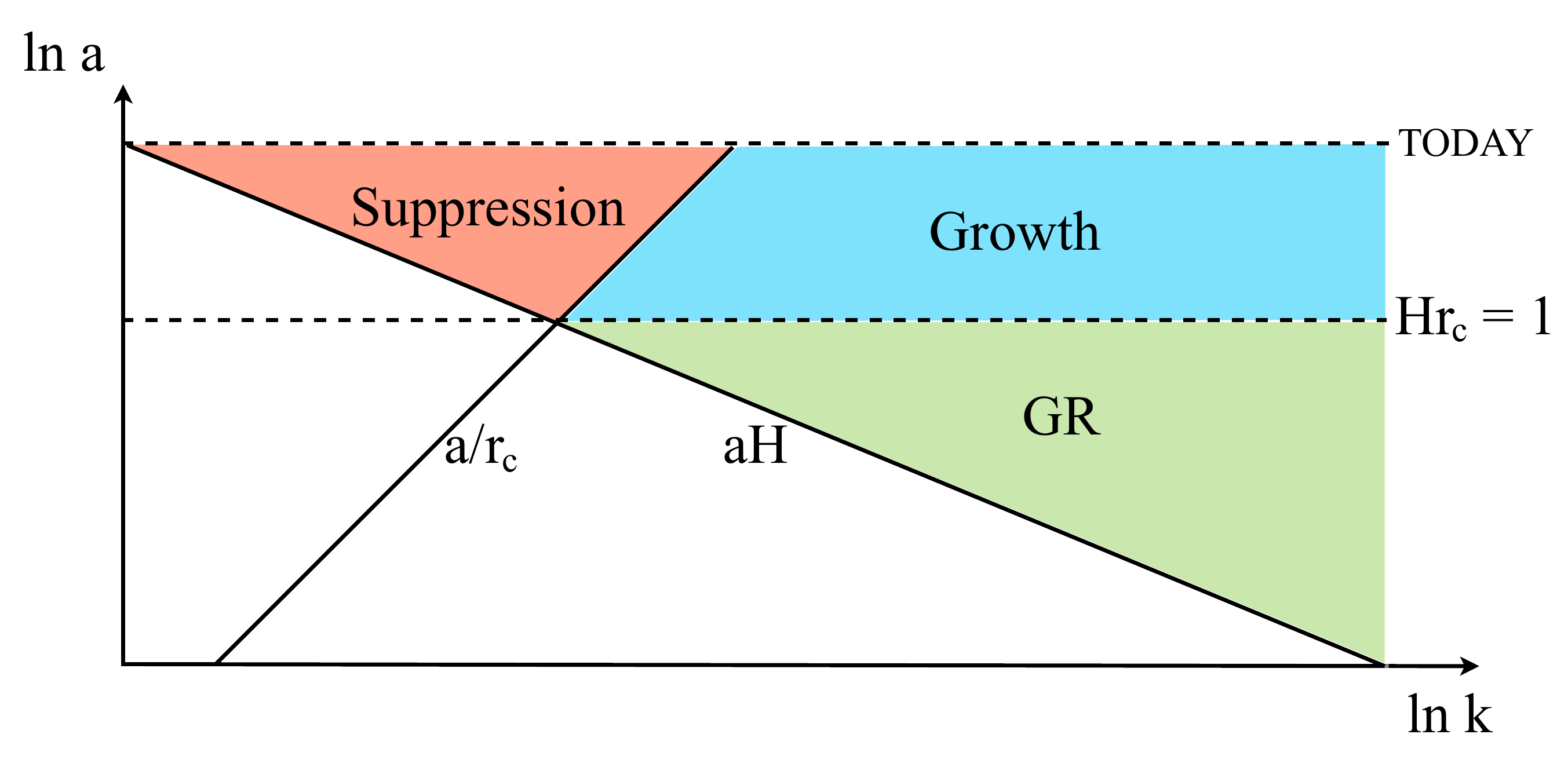}
 \caption{Different growth regimes as function of scale factor $a$ and comoving wavenumber $k$. Colored regions correspond
 to modes inside the horizon today. At early times, $H > r_c^{-1}$, growth proceeds as in general relativity (Green region); for $H < r_c^{-1}$ and on scales smaller than the graviton Compton wavelength ($k> a/r_c$), growth is enhanced thanks to the helicity-0 mode (Blue region); on large scales, $k> a/r_c$, the graviton mass suppresses growth (Red region). The Vainshtein effect
 is not included here.}
 \label{sketch} 
 \end{figure}

Before launching into the numerical analysis, we can gain analytical intuition by studying the linear regime, described by~(\ref{lin}).
The physics is illustrated in Fig.~\ref{sketch}, where the solid lines labeled by $aH$ and $r_c/a$ denote the Hubble horizon
and graviton Compton wavelength, respectively.

At early times, $H\gg r_c^{-1}$, gravity is approximately standard on all scales since $g \approx 0$. Once $H$ drops below $r_c^{-1}$, however, the scalar force mediated by the longitudinal mode kicks in, and perturbations experience enhanced growth, at least on sufficiently small scales. Thus we expect excess power on small scales compared to what is expected in $\Lambda$CDM,
at least until they reach large overdensities. 

On large scales, the graviton mass suppresses the growth of modes with $k < a/r_c$. As can be seen from Fig.~\ref{sketch},
intermediate wavelength modes ($k\lsim a/r_c$) first experience a period of enhanced growth from the longitudinal mode, followed by a period of decay; 
very long wavelength modes ($k\ll a/r_c$), on the other hand, experience only decay from horizon entry until today. 

\section{Numerical Simulations of Modified Gravity}

 Our numerical solutions were performed by modifying a publicly 
 available Particle Mesh N-body code~\cite{PMesh}; a modification of the same original code
was used in~\cite{Laszlo:2007td} for other modified gravity models.
  While many more accurate methods for computing
 the formation of large scale structure now exist for standard gravity, the particle mesh approach is actually better suited to the modification of gravity we consider here. This is because our modification,
 like the particle mesh approximation, does not accurately capture the two point interaction between
 any two of the gravitating N-bodies. On the contrary, its implicit spherical top-hat assumption only works
 over fairly long distance averaging scales, where the coarse-grained picture it relies on is a good
 approximation of reality. This is also true of the particle mesh approximation for computing
 N-body evolution. There, too, individual particles only source the gravitational potential after being 
 coarse-grained over. Recognizing this vindicates our use of the less sophisticated particle mesh approach, but will prove a difficulty as we attempt to test our model of gravity with more precision; we will discuss this issue further in Sec.~\ref{continuum}.
In short, the nonlinearities in the $\pi$ scalar field make it both  phenomenologically viable
and hard to simulate. 
  
The equations derived above are written with an implicit assumption of long wavelengths and 
Fourier transforms. 
This is what allowed us to write the d'Alembertian operator as $k^2$, for instance. Our numerical solution
of the modified Poisson equation~(\ref{modpoisson}) is of course performed using discrete Fourier transforms 
and cyclic reduction in the usual way. The effect of the modifications on the right hand side of~(\ref{modpoisson})
is to provide a nonlinearly transformed density field, with the operations all performed in real space. The result
of that transformation is then converted into Fourier space. The effect of the change on the left hand side
is even simpler. Since we have the Fourier space version of the left hand side, we simply use it
to solve for $\Phi_k$, albeit using the Green's function appropriate for the discrete Fourier transform
that appears in the code.

\subsection{The continuum limit and high density screening}
\label{continuum}
As written, our modified Poisson equation~(\ref{modpoisson}) does not have a smooth continuum 
limit: as we increase the resolution of our simulation, the nonlinearities of the equation will cause 
 a larger and larger fraction of the mass in the Universe to be screened. Fortunately, however,~(\ref{modpoisson}) is
 an approximation to an underlying equation of motion which {\it does} have a smooth limit. As described
 in~Appendix \ref{decoup},~(\ref{modpoisson}) is exact for homogeneous density, but gives a poor approximation
 in very sparse regions and whenever a significant fraction of the matter lies in high-density regions.

A moment's thought reveals that our modified Poisson equation should be applicable provided that the mass in
typical grid cells are not screened, that is, provided that the typical grid smoothing scale, $\ell_G$, is larger
the $r_\star$ corresponding to the mass in a given cell. Concretely,
let us write the density in a given cell as $\delta \rho= N M/\ell_G^3$, where $M$ is the mass in a grid cell with
mean cosmological density, and $N$ is an overdensity factor. It is easily seen that our modified Poisson equation is a
good approximation for cells satisfying 
\be
\frac{r_\star}{\ell_G} \simeq \frac{N^{1/3}}{15} \( \frac{r_c}{400 \mbox{ Mpc}} \)^{2/3} \; \lsim\; 1\,.
\ee
In other words, any grid cell that has an overdensity of less than $4\times10^{3}$ compared
with the average density today will not experience substantial
screening. This is consistent with Fig.~\ref{antiV}, where the suppression of the extra
force is significant only for very large overdensities. Note that our criterion is independent of
the number of particles in the simulation because we have averaged over each grid cell.
This is a good approximation when $N_{\rm g} \lesssim 2 N_{\rm p}$, where $N_{\rm g}$ and $N_{\rm p}$ denote the number of grid cells and particles, respectively,
since the Cloud-in-Cell density assignment scheme used in our code smears individual particles over several grid cells. 

Of course the above criterion is harder to maintain at early times, since the density is larger. 
But this is neither a surprise nor a concern, since extreme overdensities on the length 
scales of interest are rare until late times, cosmologically speaking. 
Moreover, at early times any enhanced force effect is quenched by the cosmological screening through
the parameter $\beta$.

Even in the regime where~(\ref{modpoisson}) is valid, we can check whether a more dense
grid leads to a systematic reduction in structure growth over the range of scales
where we trust our simulations. The results of this test, shown in Fig.~\ref{res} and
discussed in detail in Sec.~\ref{avg}, show no such systematic effect. 

%
%The upshot of this is that the high-density suppression form of the modified Poisson equation
%plays, in fact, rather a small role in cosmological calculations. This is good in that it gives 
%us confidence that our simulations are approximating the true behavior of the underlying theory 
%well for the questions that are our focus. However, it also points up the necessity for finding 
%a better approximation scheme if we wish to consider either larger values of $r_c$ or
%to study phenomena, like clusters, that are dominated by high-overdensity behaviors. 

% The particle mesh Fourier
% tranform solver, however, includes all frequency modes that the computational grid can resolve,
% including relatively large frequency modes. To apply our formulae, we have made the simple
% substitution:
% \be
% k_i \rightarrow \sin{k_i}\;; \quad \quad k_i = 2 \pi/ L,
% \ee
%where $L$ is the box size in code units. This is only relevant for the graviton propagator, the left-hand side
%of Eqn. \ref{modpoisson}, though. The right-hand side is a nonlinear transformation of the density field, so we apply
%it to the discretized density field on the mesh first, then Fourier transform the resultant effective density field to obtain
%the effective Newtonian potential. 

\begin{figure} %  figure placement: here, top, bottom, or page
%\begin{figure}[htbp] %  figure placement: here, top, bottom, or page
   \centering
   \includegraphics[width=0.5\textwidth]{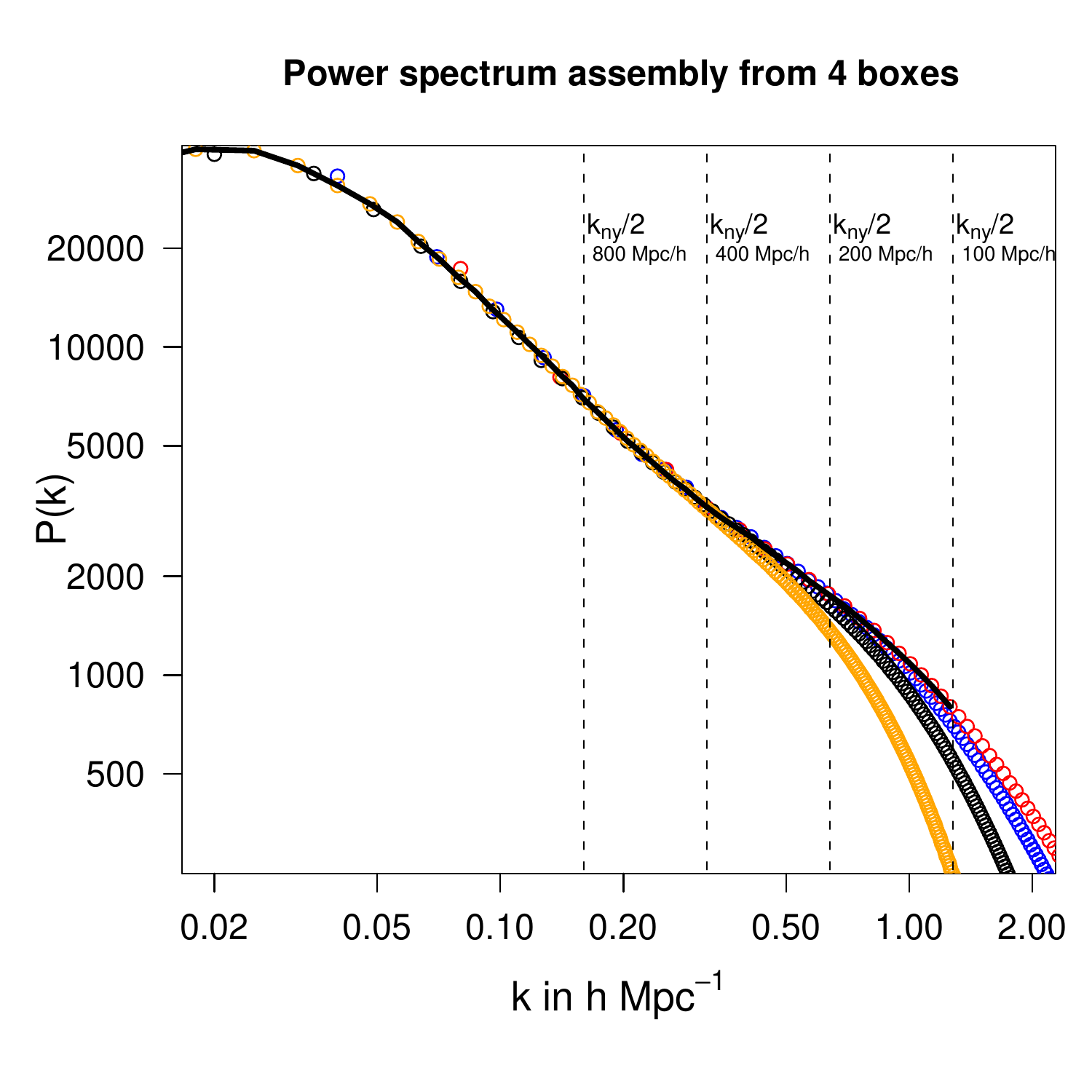} 
   \caption{This plot demonstrates how a full range power spectrum is assembled from our four 
   overlapping simulation boxes. The vertical dashed lines indicate the half Nyquist frequencies for the
   four boxes. The orange, black, blue, and red circles represent the averaged power spectra for the
   800, 400, 200, and 100 $h^{-1}$Mpc boxes, respectively. The simulation pictured is that for $r_c=500$ Mpc in the 
   degravitation model ($\alpha = 0$).}
   \label{build}
\end{figure}

\section{Power Spectrum}
\label{powspec}

\begin{figure}[htbp] %  figure placement: here, top, bottom, or page
   \centering
   \includegraphics[width=0.5\textwidth]{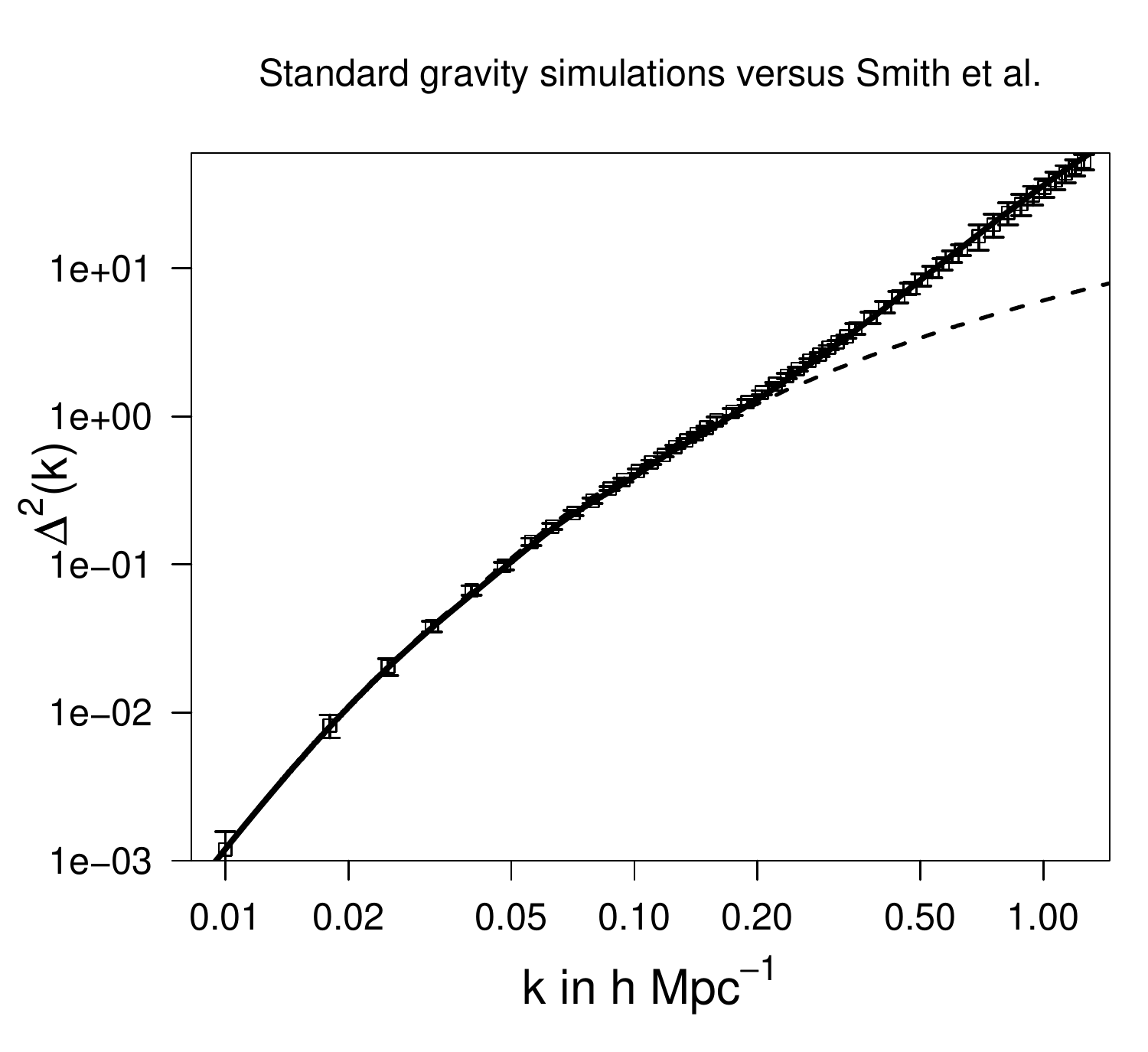} 
   \caption{For standard gravity, this shows a comparison of $\Delta^2(k) = k^3P(k)/2\pi^3$ between our power spectrum
   and that obtained through the Smith {\it et al.} non-linear fit~\cite{Smith:2002dz}, which is represented by the solid black line.
  The dashed line is the linear prediction.}
      \label{smithcomp}
\end{figure}

We perform a variety of runs --- approximately 40 replicates for each parameter combination ---
each with $512^3$ grid points and $256^3$ particles. The model parameter values tested in the runs
are $r_c = 300$ and $500$ Mpc, with $\alpha=0$ and $0.5$. (The values of  $r_c$ quoted here are all in physical Mpc, as opposed to $h^{-1}$Mpc; all other quantities throughout the paper are expressed in the usual $h^{-1}$Mpc units.)
We also perform a set of runs with identical initial conditions assuming standard gravity. 

Each parameter combination is run in boxes of size $800$, $400$, $200$, and $100\, h^{-1}$Mpc. 
For each of our four simulation boxes, we can compute a power spectrum. What we want, though, is a combined power      
spectrum that includes information over a wide range of scales. The power spectrum for a box of a given size $L$ and         
containing $N_p$ particles, is accurate down to a length scale $2L /  N_p$, or, equivalently, a frequency scale $k_{\rm Ny} =       
N_p/2L$ --- the half Nyquist frequency. We therefore join our power spectra together in the following way.          
Starting with the largest box, we keep the power spectrum for modes between the smallest $k$-mode that the box can resolve     
({\it e.g.}, 0.01~${\rm h\, Mpc}^{-1}$ for the 800~$h^{-1}$Mpc box) and the half Nyquist frequency. We then continue the power spectrum by          
taking the next larger frequency mode from the next smaller box. That is, the power at $k_{\rm Ny}$ for the 800~$h^{-1}$Mpc box     
is followed in the composite power spectrum by a power value determined from the 400~$h^{-1}$Mpc box. We then keep   
all of the power spectrum points from the 400~$h^{-1}$Mpc box between that first frequency and the 400~$h^{-1}$Mpc half Nyquist frequency.
The procedure is repeated to include points from the 200 and 100~$h^{-1}$Mpc boxes. 
The power spectrum determined from our simulations thus ends at the half Nyquist frequency of the 100~$h^{-1}$Mpc box.  
The constituent parts and the resulting power spectra are illustrated in Fig.~\ref{build} for degravitation model ($\alpha = 0$) with $r_c = 500\;{\rm Mpc}$. The transition between simulation boxes is marked by jumps in scatter among runs and hence variance. This 
is a result of our decision not to fix the overall normalization of power within the smaller boxes, but to let it be randomly sampled.
The modified simulations show larger scatter due to their enhanced growth, which exaggerates the differences between different
initial conditions among small box realizations.

For standard gravity, we can test the accuracy of our simulations by comparing our power spectrum with
the non-linear fitting algorithm of Smith {\it et al.}~\cite{Smith:2002dz}. The result, displayed in Fig.~\ref{smithcomp},
shows very good agreement.

\bwt

\begin{figure}[h!] %  figure placement: here, top, bottom, or page
   \centering
   \subfigure[ $\;\alpha=0$]
   {
   	\label{pkzero}
	\includegraphics[width=0.45\textwidth]{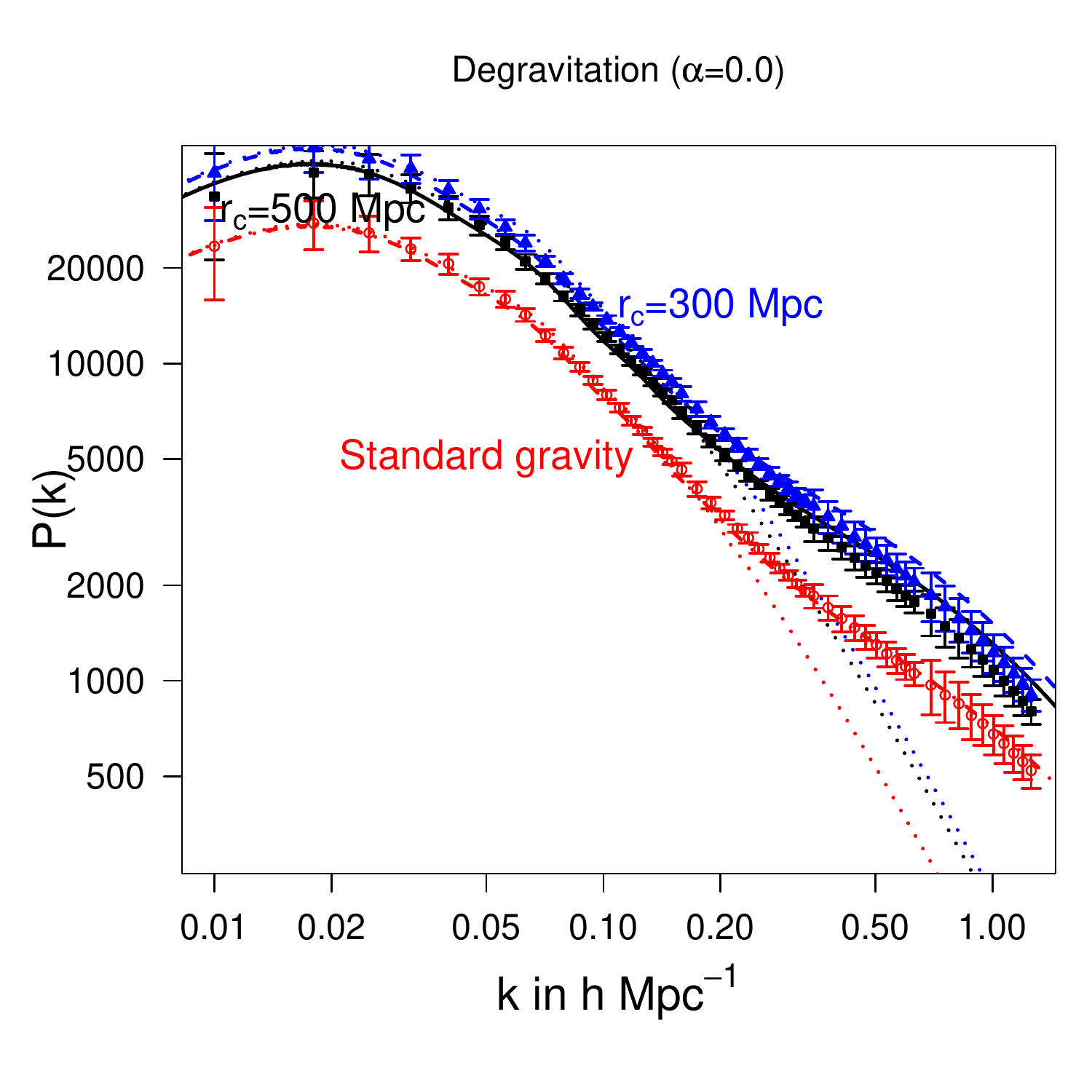}
	}
	   \subfigure[ $\;\alpha=0.5$]
   {
   	\label{pkhalf}
	\includegraphics[width=0.45\textwidth]{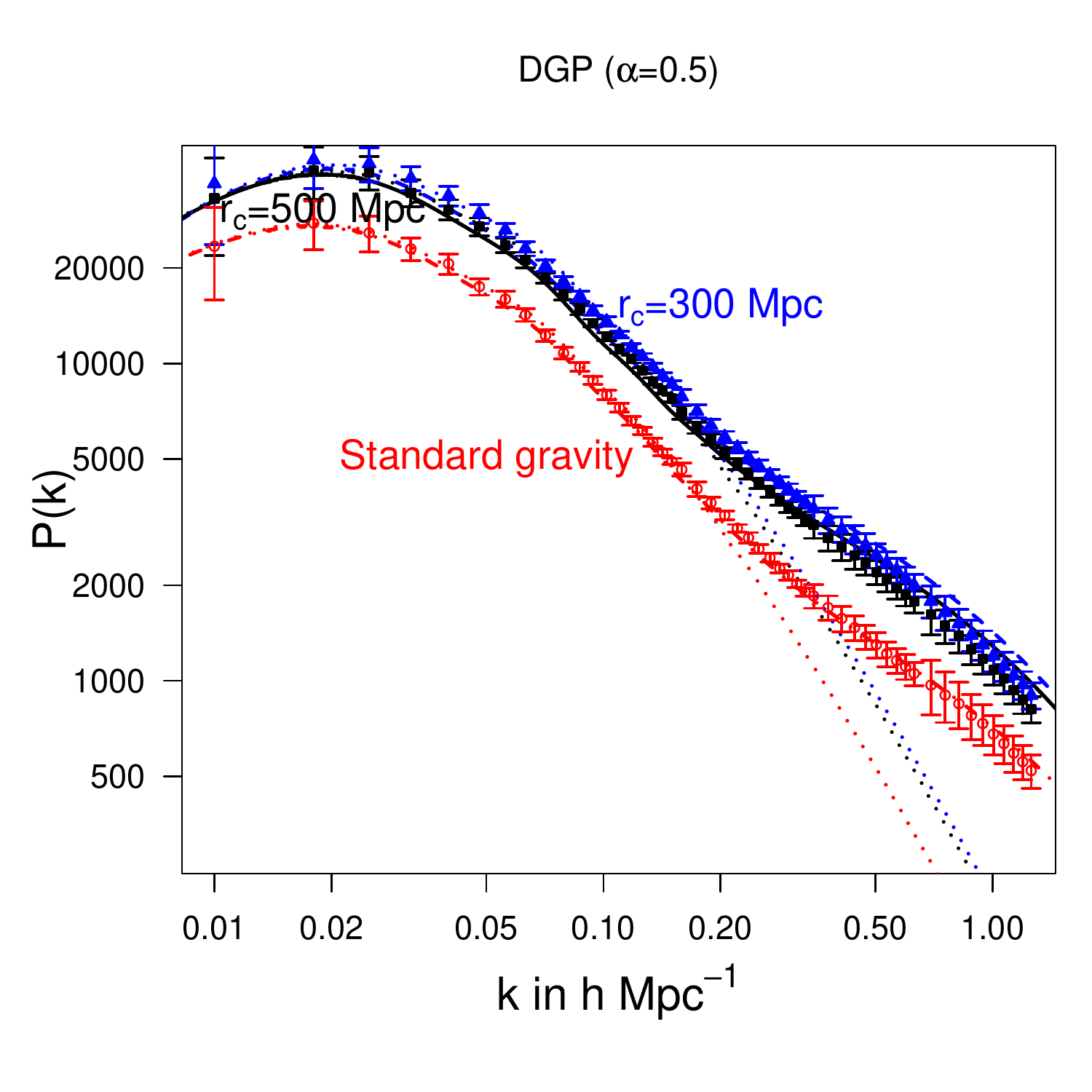}
	}
   \caption{The CDM power spectrum from our N-body codes, assembled from the four simulation boxes
   in the way described in the main text. The blue (black) points are for $r_c=300$   (500) Mpc. The red points are for standard gravity. The error bars represent the variance among 
   numerical runs. Each has been reconstructed by combining results from four simulation boxes,
   and averaged over a number of realizations. The dotted lines show the results from solving
   the linear perturbation equations; the solid (dashed, dash-dotted) lines are the results from the Smith fitting procedure
   for $r_c= 500$ Mpc (300 Mpc, standard gravity).}
   \label{pkvsmith}
\end{figure}

\begin{figure}[h!] %  figure placement: here, top, bottom, or page
   \centering
   \subfigure[ $\;\alpha=0$]
   {
   	\label{kpkzero}
	\includegraphics[width=0.45\textwidth]{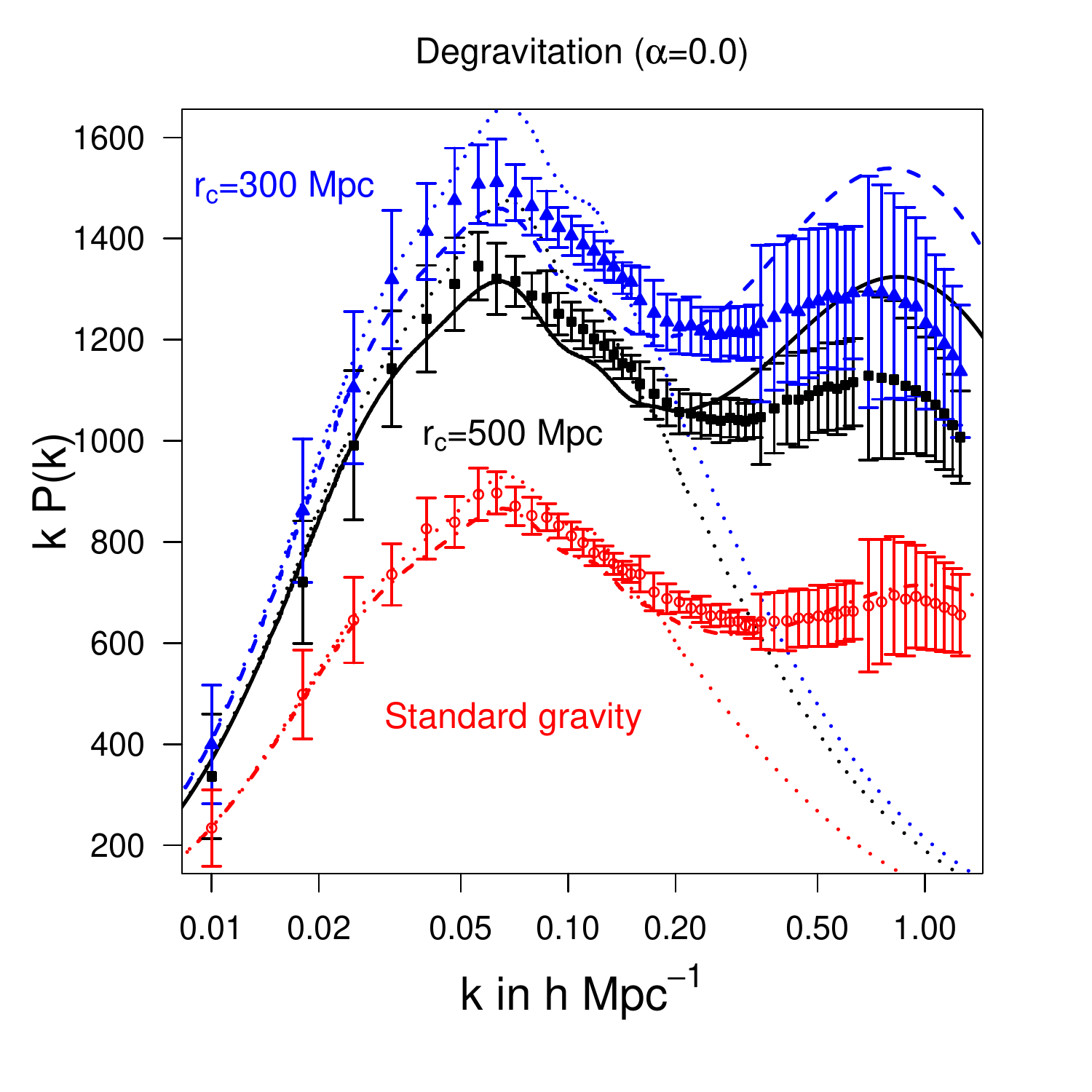}
	}
	   \subfigure[ $\;\alpha=0.5$]
   {
   	\label{kpkhalf}
	\includegraphics[width=0.45\textwidth]{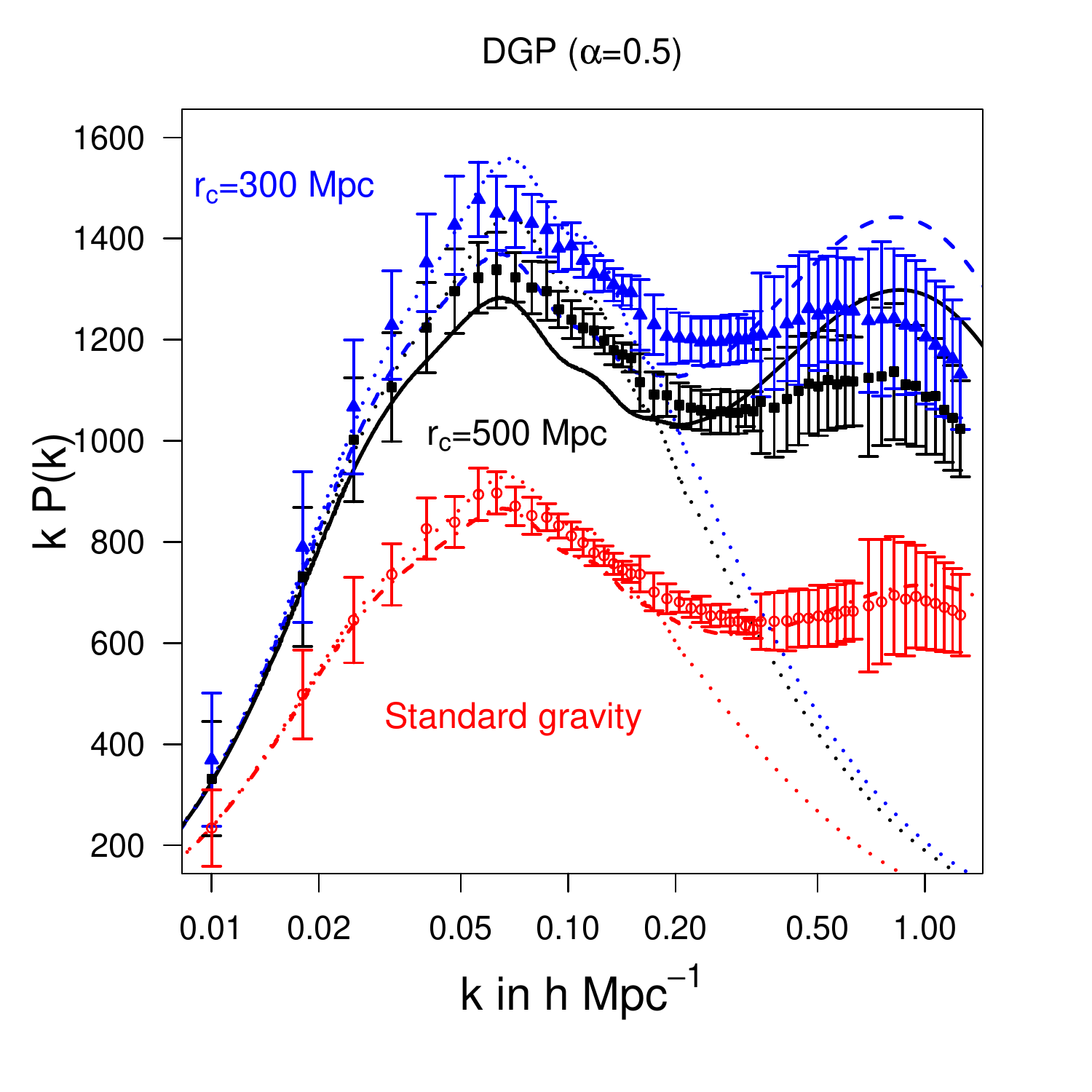}
	}
   \caption{Plot of $kP(k)$ for our simulations. The convention for curves and data points is identical to that of Fig.~\ref{pkvsmith}.}
   \label{kpkvsmith}
\end{figure}

\begin{figure}[h!] %  figure placement: here, top, bottom, or page
   \centering
   \subfigure[ $\;\alpha=0$]
   {
   	\label{kpkzerob}
	\includegraphics[width=0.45\textwidth]{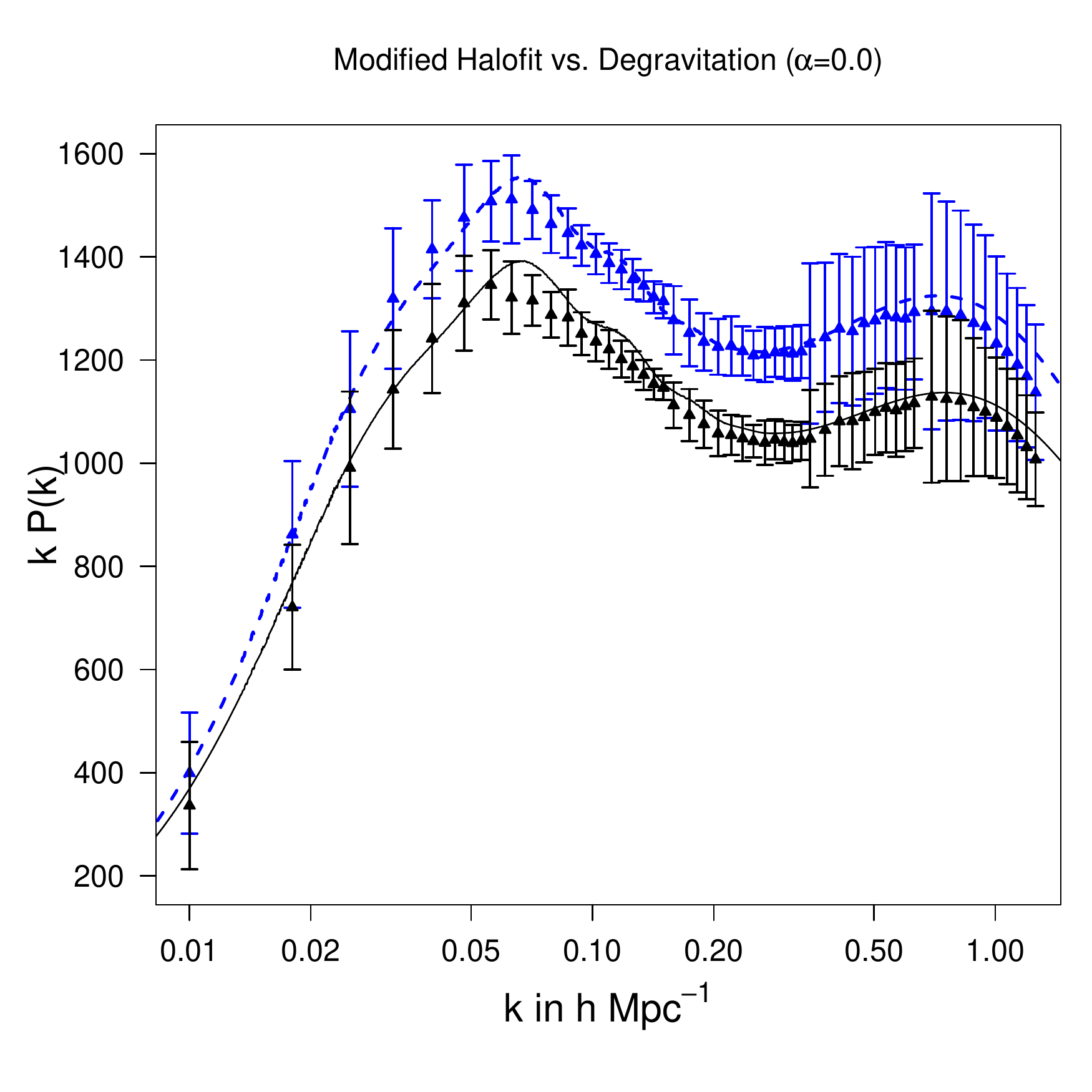}
	}
	   \subfigure[ $\;\alpha=0.5$]
   {
   	\label{kpkhalfb}
	\includegraphics[width=0.45\textwidth]{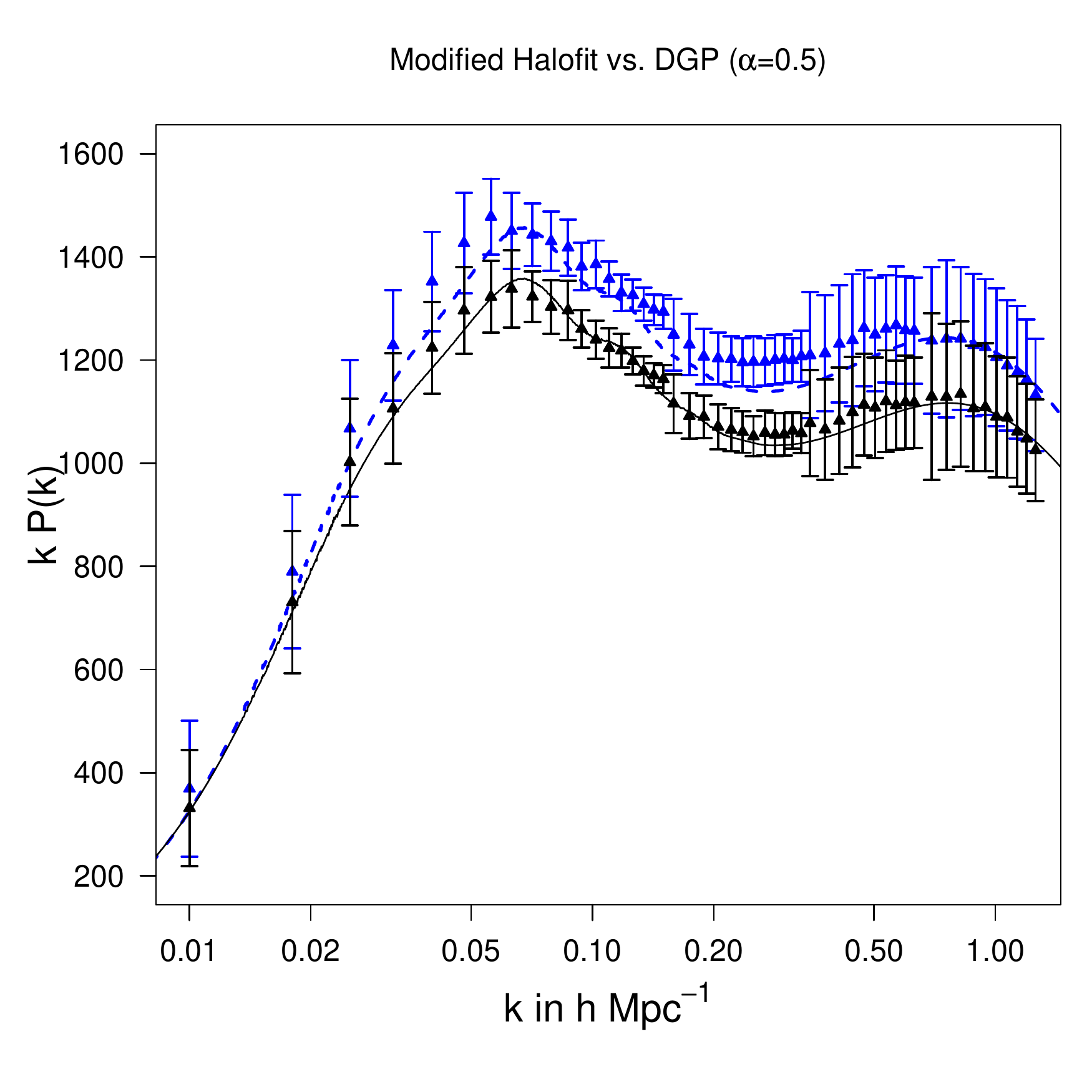}
	}
   \caption{Comparison of our modified non-linear fitting algorithm with power spectra from our DGP (left panel) simulations 
   and degravitation/cascading (right panel) simulations, each with $r_c = 300$ (blue dashed curve) and $500$~Mpc (black solid curve). Again, the error bars represent the variance among numerical runs. The modified halofit parameters are common
   to both $\alpha$ choices and are listed in Table~\ref{smithparams} of Appendix \ref{smithfit}.}
    \label{modsmith}
\end{figure}

\ewt

\begin{figure}[h] %  figure placement: here, top, bottom, or page
   \centering
   \includegraphics[width=0.5\textwidth]{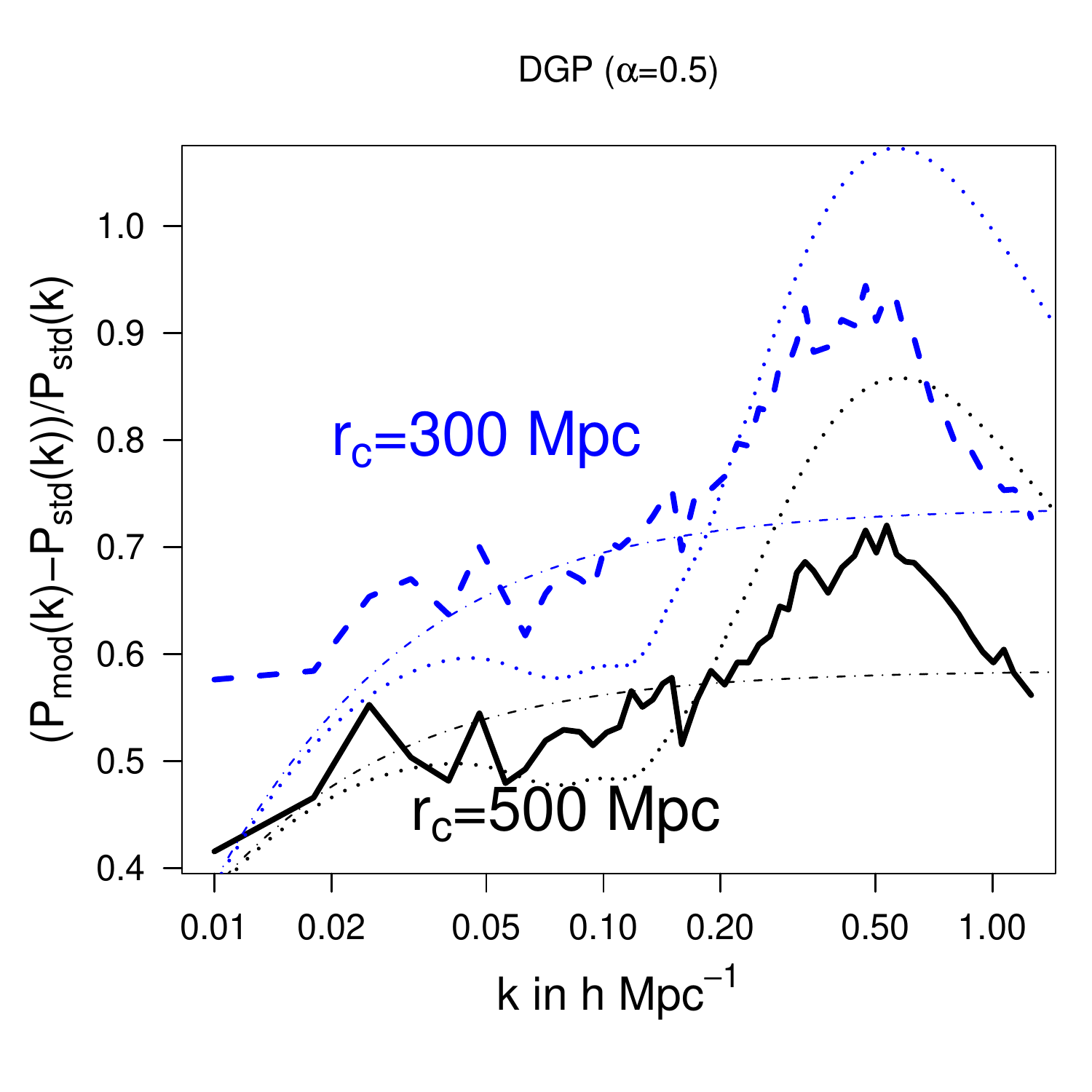} 
   \caption{The same as Fig.~\ref{fracdiff} for the DGP parameter choice, $\alpha=0.5$; $r_c=300$ Mpc is represented by the blue dashed line and $r_c = 500$ Mpc by the black solid line. The dash-dotted lines are the expected difference from linear perturbation theory. The dotted lines are the expected difference assuming the Smith {\it et al.} procedure~\cite{Smith:2002dz} for including nonlinear effects. }
      \label{half fracdiff}
\end{figure}

\begin{figure}[htbp] %  figure placement: here, top, bottom, or page
   \centering
   \includegraphics[width=0.5\textwidth]{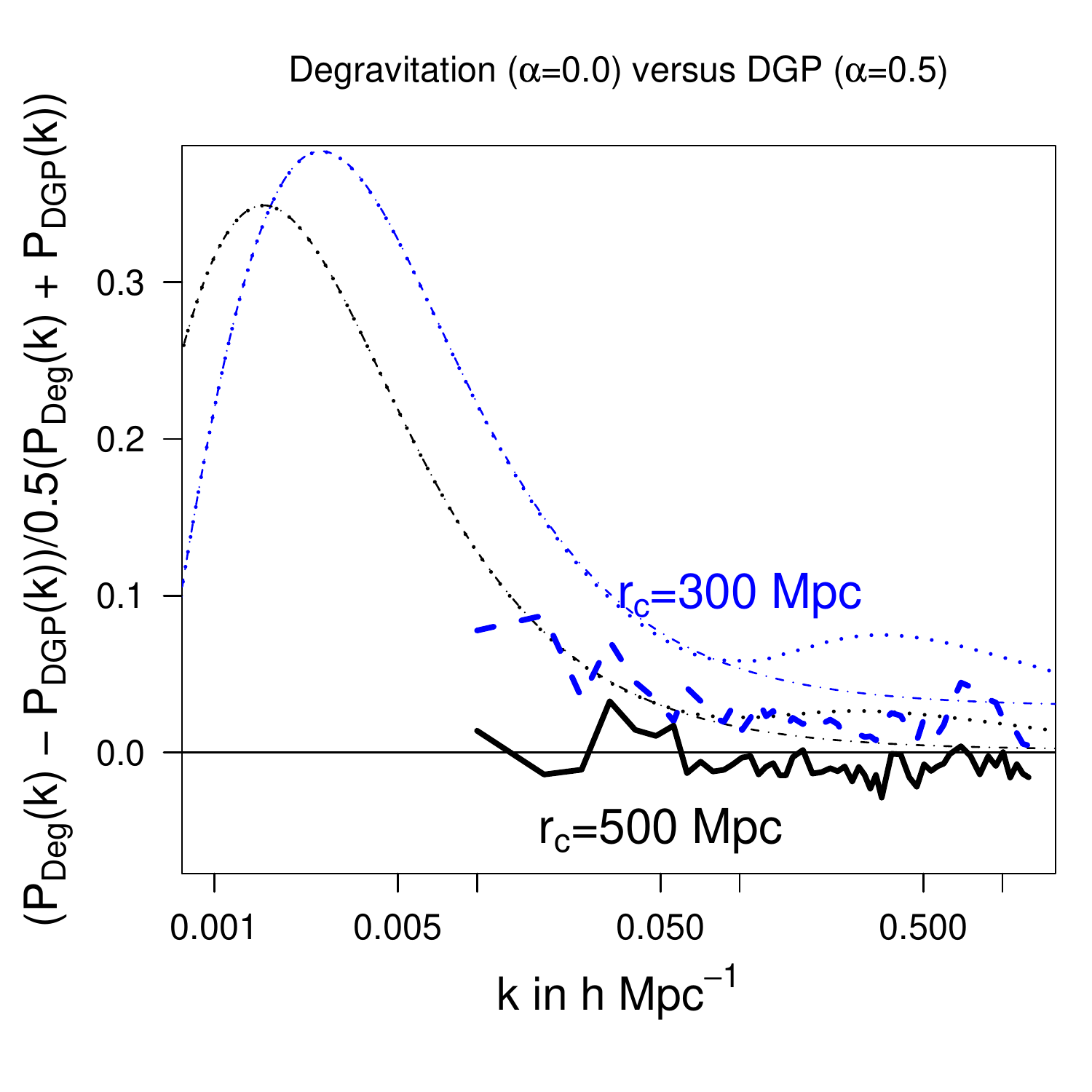} 
   \caption{Comparison of the simulations for DGP ($\alpha=0.5$) and degravitation/cascading models ($\alpha =0.0$).
   The linear theory and nonlinearly corrected linear predictions are plotted as dash-dotted
   and dotted lines, respectively, for each value of $r_c = 300\; (500)$~Mpc in blue (black).
   The differences are small on the scales of the simulations, but become significant on larger scales.}
   \label{dgpvdeg}
\end{figure}

\begin{figure}[htbp] %  figure placement: here, top, bottom, or page
   \centering
   \includegraphics[width=0.5\textwidth]{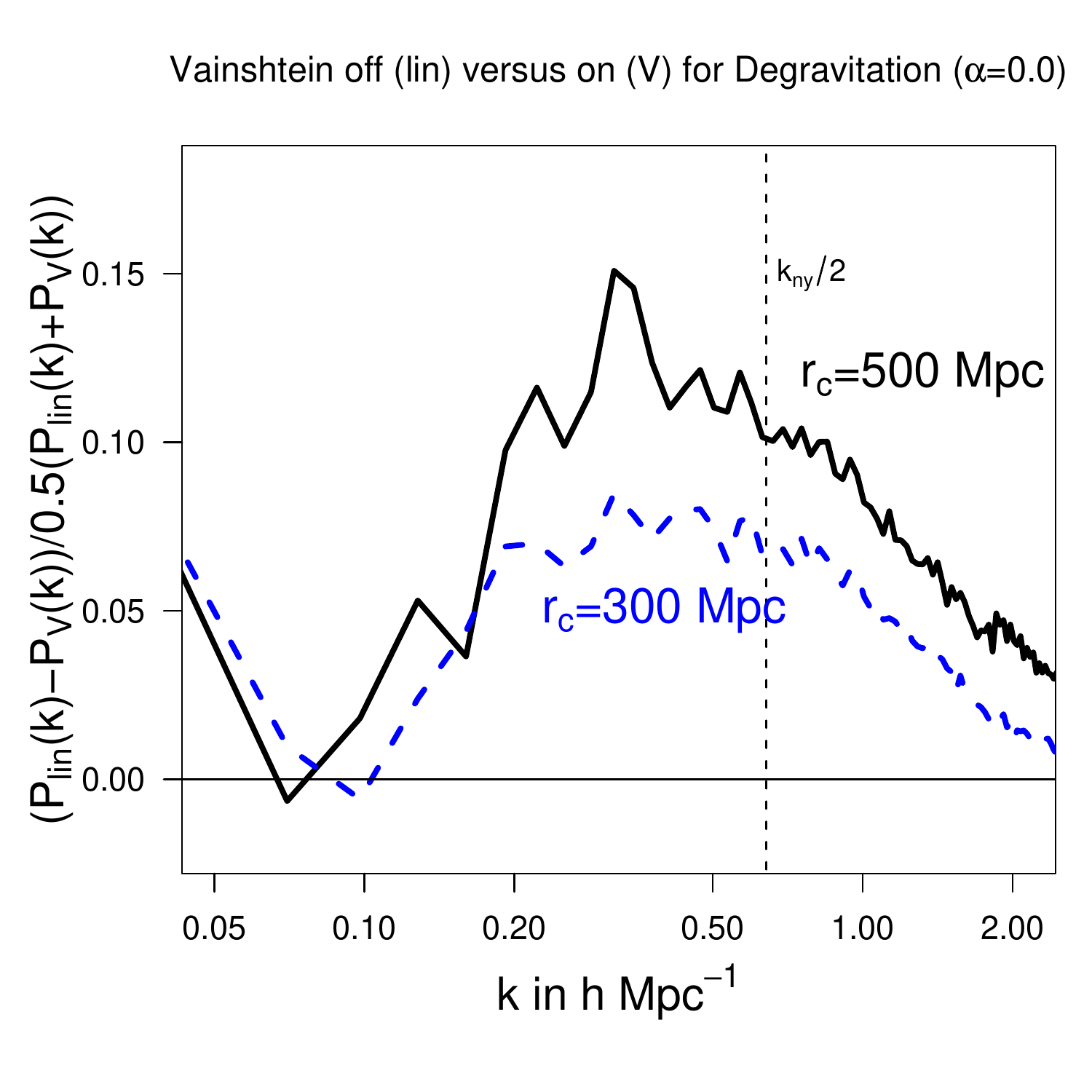} 
   \caption{A clear illustration of the local Vainshtein screening effect, due to large overdensities. 
    This plot was made using a 200 Mpc/h box for a degravitation/cascading
    model ($\alpha=0$), with $r_c=300$ Mpc plotted as a blue dashed line, 500 Mpc plotted
    as a black solid line. The black vertical dashed line shows the
   half Nyquist frequency for this box. The local Vainshtein effect leads to somewhat less development
   of power in the trustworthy portion of the power spectrum, as expected. Since the Vainshtein suppression
   is only effective for high overdensites ($\delta \gtrsim 100$), however, its
   strongest manifestation is not captured by the simulations performed here.}
   \label{chameleon}
\end{figure}

\begin{figure} %  figure placement: here, top, bottom, or page
%\begin{figure}[htbp] %  figure placement: here, top, bottom, or page
   \centering
   \includegraphics[width=0.5\textwidth]{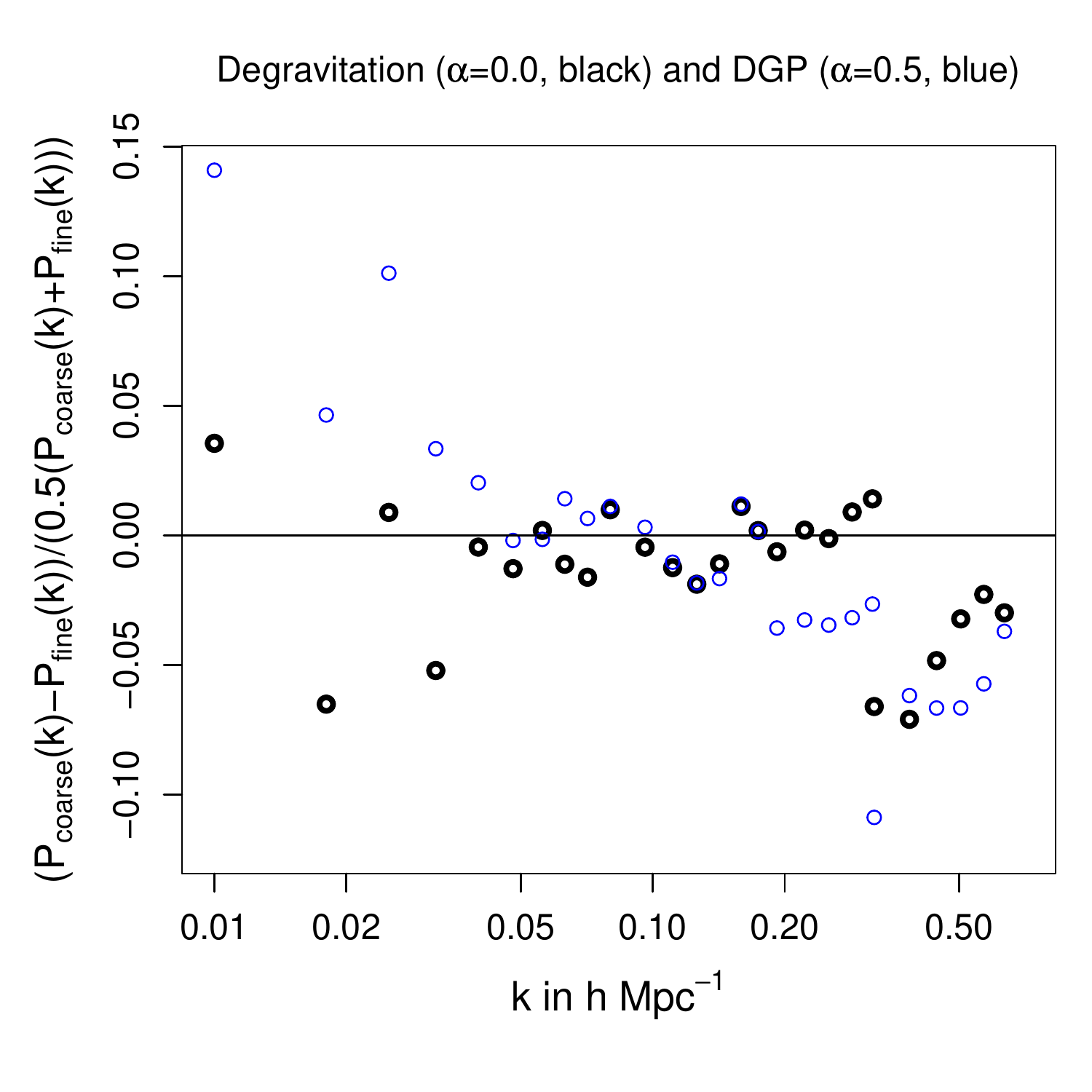} 
   \caption{This plot demonstrates the dependence of our results on box resolution by showing the fractional
   difference between the full resolution runs and a set of simulations done with ``half" resolution in both
   grid size and particle number (ie. a 256$^3$ grid with 128$^3$ particles).  The black circles shows the
   results for a $r_c = 300$ Mpc cosmology for a cascading / degravitation model, the blue circles the results for a DGP model.
   Note that the results for the finer resolution have more power on small scales than the coarser grid shows, even though
   we have restricted this plot to those scales when both meshes should give valid results. This gives further
   evidence that higher grid resolutions are not causing significant high density screening, as argued in \S \ref{continuum}. }
   \label{res}
\end{figure}

\subsection{Comparison with non-linear fitting formulae}

Figures~\ref{pkvsmith} and~\ref{kpkvsmith} compare power spectra for DGP and degravitation models with standard gravity.
The quantity plotted in Fig.~\ref{kpkvsmith} is the power spectrum multiplied by the wavenumber, $kP(k)$, to highlight features
on non-linear scales. As anticipated, the extra scalar attraction due to $\pi$ enhances gravitational clustering compared to $\Lambda$CDM cosmology.
Power is increased substantially on a wide range of scales, and even more so on scales that have undergone nonlinear evolution.
The relative enhancement compared to standard gravity is quantified in Figs.~\ref{fracdiff} and~\ref{half fracdiff} for degravitation and DGP
models, respectively.

Also plotted in Figs.~\ref{pkvsmith} and~\ref{kpkvsmith} are the results of the Smith {\it et al.} fitting algorithm~\cite{Smith:2002dz} in each case.
We see that the fitting approach agrees very well with our simulations over the entire range of $k$ for standard gravity,
and for the linear and weakly non-linear scales in the modified gravity models. On more strongly nonlinear scales ($k\gsim 0.2\; h\,\rm{Mpc}^{-1}$), however, 
the Smith {\it et al.} algorithm heavily overpredicts power relative to what we find in our simulations of modified gravity. This is most easily seen in
the high-$k$ ``bump" in Fig.~\ref{kpkvsmith}, spanning the range $0.2 \lesssim k \lesssim 1.2\; h\, {\rm Mpc}^{-1}$. The modified gravity ``bump" is more pronounced than in standard gravity --- as can be seen in Fig.~\ref{vdata} --- but not as large as what the Smith {\it et al.} algorithm predicts. 

This mismatch on small scales is attributable to the Vainshtein mechanism present in our equations, which is not captured by the Smith {\it et al.} procedure.
In modified gravity, sufficiently small-scale structures enter the non-linear regime early enough that the $\pi$-mediated force is negligible ($g\approx 0$) at that time.
Subsequently, as they become increasingly non-linear, such perturbations simply cease to feel any influence of $\pi$. Sufficiently small-scale modes thus never
experience the effects of modified gravity. Intermediate-scale perturbations, on the other hand, experience a phase of $\pi$-enhanced growth before
going non-linear.  

This leads to a scale-dependent pattern of structure growth that is different from that expected from general relativity. The Smith {\it et al.} algorithm,
on the other hand, was calibrated against simulations of standard gravity that were --- apart from the horizon scale --- genuinely scale-free. The algorithm
simply scales the non-linear power according to the linear prediction and, as a result, overpredicts small-scale power for our model. This mismatch is also present 
in simulations of $f(R)$ gravity~\cite{marcos}, in that case due to the chameleon mechanism.

Remarkably, it is straightforward to recalibrate the Smith {\it et al.} formulae to capture the distinctive development of nonlinear structure in our simulations. We show the results in Fig.~\ref{modsmith}. To achieve this, we simply rescaled most of the
Halofit algorithm's numerical coefficients until we found a set that fit our simulations. The parameter adjustments necessary
are listed in Table~\ref{smithparams} of Appendix~\ref{smithfit}. Because the difference between the DGP and cascading/degravitation simulations is small over the 
range of our simulations --- and expected to grow smaller for larger values of $r_c$ --- we have provided a single set of fitting
parameters. For small values of $r_c$ these two models become increasingly different, so the compromise parameters fit less well. 
The better fitting parameters are mentioned in the Appendix. A code for calculating these fitting formulae will be made publicly available~\cite{code}.

This result at first appears to disagree with another study, \cite{Laszlo:2007td}, that found good agreement between Halofit and 
a variety of weak modifications to gravity, including a phenomenological model that mimicked the DGP scalar
that is also our focus. Their work, however, focused on the case where the change in the scalar force was explicitly correlated
to an alteration of the expansion history, and where it mediated a repulsive force. This constrained their modification of the growth rate to be relatively small and to work against, rather than in concert with, the Newtonian force.
For very large values of $r_c$, our model also recovers standard gravity smoothly, and for the large values of
$r_c$ considered in \cite{Laszlo:2007td}, we would similarly find much better agreement with the Halofitting algorithm.

%To summarize, our models have power spectra with a large enhancement of linear growth and a suppressed enhancement to nonlinear
%processing. As our understanding of these models matures, it would be useful to recalibrate the Smith {\it et al.} formulae to capture the
%distinctive development of nonlinear structure in them, as such a fitting algorithm would permit a much faster exploration of cosmological parameter
%space.

\subsection{Dependence upon model parameters}

Another important question is how the growth history depends on $\alpha$.
Figure~\ref{dgpvdeg} shows the relative difference between power spectra for
degravitation/cascading ($\alpha= 0$) and DGP ($\alpha =1/2$) model. 
The difference between them are only a few percent over the scales probed by the simulations and
are thus smaller than the variance in the simulations, which are suppressed in the plot. This can be
seen most clearly in the results from the linearized perturbation theory results (dash-dotted lines). The 
one clear difference that is evident both in the linearized results and in the simulations is that 
the difference between the two values of $\alpha$ is more pronounced for smaller $r_c$,
which is sensible. 

The dependence of our results on the graviton Compton wavelength is simpler. 
Smaller $r_c$ leads to more power on intermediate scales, but reaches
the ``cut off" beyond which gravity ceases to operate at larger $k$. 
As is implicit in Fig.~\ref{vdata}, over the scales we simulate and for the
relatively small difference in values of $r_c$ we consider, a simple linear rescaling
is all that is necessary to account for the difference in $r_c$. 
This rescaling can be approximated analytically using Eqn. \ref{modpoisson}
and the equations governing linear perturbation growth (see, e.g.,~\cite{niayeshghazal}).
The enhancement over standard gravity scales approximately as $(H_o r_c)^{0.25}$,
so the linear rescaling necessary to go from one $r_c$ to another is approximately
$(r_c[1]/r_c[2])^{0.25}$. This rescaling matches our simulations quite well  (see, 
for instance, Fig.~\ref{fracdiff}, where a $(5/3)^{0.25}$ would put the two lines
right on top of one another). This rescaling will eventually break down at small $k$, however,
as we can see in the linear solution for small $k$ in Fig.~\ref{dgpvdeg}. It will 
also fail, along with the rest of our approximations, as we go to high $k$ where Vainshtein
screening becomes systematically large. 

Besides further elucidating how our phenomenological model depends
on its input parameters, this result demonstrates that it would be possible in principle
for cosmological observations to discriminate between different kinds of higher dimensional
gravity theories, since the differences between their predictions can be quite pronounced, especially
on very large length scales. 

\subsection{Vainshtein mechanism and large scale structure}

The Vainshtein screening effect in our class of models arises in two ways: $i)$ through a {\it global} Vainstein effect
due to the cosmological background density, encoded in the function $g$ defined in~(\ref{g}); $ii)$ through a {\it local} Vainshtein
effect from overdensities, encoded in the non-linear $\epsilon$-dependence in~(\ref{modpoisson}) --- see Fig.~\ref{antiV}. To disentangle these two effects,
we performed an additional set of simulations  where, instead of using~(\ref{modpoisson}) as the modified Poisson equation, we used its linearized form given in~(\ref{lin}). The latter does not include the Vainshtein screening from local overdensities, only from the cosmological background.
In other words, this comparison allows us to isolate the effects from local Vainshtein screening.

The results of this comparison in a representative simulation box are shown in Fig.~\ref{chameleon}. The results for each simulation box ---  and the combined power spectra --- were similar, but focusing on a single box allows us to see the changes due to the local Vainshtein effect.

The main take-away message from this figure is that the power is greater without the local Vainshtein effect, as expected intuitively.
More interestingly, the power excess is larger for $r_c = 500$~Mpc then for 300~Mpc. Indeed, for larger $r_c$,
non-linearities in $\pi$ are more easily triggered, hence the local Vainshtein screening is more effective. Quantitatively, local screening
requires a large overdensity in units of $r_c$, that is, $\epsilon\gg 1$. And from~(\ref{eps}), $\epsilon\gg 1$ is easier to achieve for larger $r_c$. 
(Of course, for much larger values of $r_c$, such that $H_0r_c\gg 1$, the global Vainshtein screening would still be effective today, and
the local Vainshtein effect would therefore be irrelevant.)

Going back to Fig.~\ref{chameleon}, the reason that the difference between the two simulations goes to zero past the half Nyquist frequency is because the simulations cease to give a valid determination of the power spectrum beyond that point. Intuitively, we expect that the reason why these highest 
frequency modes have more power in the linear simulation is because of the enhanced kinetic energy of the particles in
that simulation: since the mesh cannot resolve the inter-particle potential on small scales, we can see that in simulations with stronger 
gravity that the two particle correlation falls off more quickly than in simulations where gravity is weaker on smaller length scales. 

\subsection{Sensitivity to averaging}
\label{avg}

As discussed in Secs.~\ref{CP} and \ref{continuum}, our modified Poisson equation~(\ref{modpoisson}) requires averaging the density field.
Since the equation depends non-linearly on $\delta$, one might have expected our results to depend on the
coarse-graining scale. As discussed in \S \ref{continuum}, the relevant scale is set by the grid resolution. Therefore, we can quantify the dependence on
the scale of averaging by comparing simulations of different resolution. Figure~\ref{res} shows that degrading the resolution
of runs by a half, both in grid size and in particle number, results in a $\lesssim 10$\% change in power. Most importantly, 
the difference is smallest over the intermediate length-scales where our simulations are most accurate. This 
demonstrates that the short distance, high-density screening discussed in \S~\ref{continuum} is not propagating its effects to these
length scales. Indeed, we even see that the finer grid --- which resolves more high density structures in a universe with hierarchical clustering ---
has {\it more} power on small scales than the coarse grid, whereas the effect of spurious screening would be to decrease the power
present at higher resolution. 

\subsection{Confronting observations}

The relatively poor fit of our simulation power spectra to the SDSS data shown in Fig. \ref{vdata} 
should not cause alarm. This is due to several factors, most obviously that our chosen cosmological parameters
were chosen for model comparison purposes, not to fit these data. Also, our simulations, which contain only dark matter,
will necessarily match real data poorly. 

With regard to data, the most obvious difference between the modified and standard models is that 
the bias needed for the modified gravity models to fit these data is very different
from the typical bias parameter --- 56\% (67\%) for 300 (500) Mpc relative to that needed for the
standard gravity results. This is a natural consequence of our model's enhanced gravitational strength; its
consequences bear further study. 
At the present, the theory of bias in modified gravity models has yet to be worked out.
The study bias in these models is complicated by the fact that they introduce scale dependence into the
growth factor.  There are also indications
in related work~\cite{niayeshghazal} that a more complete phenomenology of a massive / resonance
graviton model may require a more subtle treatment of super-horizon evolution. 
Both of these effects will introduce scale dependence into the bias.

A more pressing concern for these models comes from recent precision weak lensing measurements~\cite{Fu:2007qq}.
These observations observe a power spectrum normalization of $\sigma_8 (\Omega_{\rm m}/0.25)^{0.64}=0.837\pm0.084$ on $\theta > 85'$ angular scales,
in good agreement with the WMAP measurement~\cite{wmap5}: $\sigma_8=0.80\pm0.04$.
This would appear at first to rule out a significant boost in power spectrum amplitude from  modified gravity.
However, lensing data are not so easy to interpret in the context of our modified gravity model, and do not provide a
strong constraint on it as yet. 

A simple way to see this is to compare with the results in~\cite{Dore:2007jh}, which studies the consequences for weak lensing of
adding a mass term in the Poisson equation. This is of course very different from the density-dependent modifications considered here.
Nevertheless, as can be seen from the growth factor shown in their Fig.~1, our class of models is mimicked reasonably well by their
Yukawa potential with $\alpha_{\rm Dore} = -0.2$ and inverse mass $\simeq 0.1-1$~Mpc (similar to our $r_\star$).
Using these values, we see that, according to their Fig. 2, this model is as yet in no conflict with the weak lensing data. Furthermore,
because of the bias issues discussed above, the comparison with the Sloan Luminous Red Galaxies also shown in their Fig.~2
should be ignored.

Another subtlety to keep in mind is that weak lensing observations are sensitive to a different gravitational potential, usually denoted by $\Phi_-$.
This quantity is related to the Newtonian potential by
\be
\Phi_- = -\frac{\Psi}{1-g}\,.
\label{phi-}
\ee
We can understand this result by noting that, unlike non-relativistic particles, photons do not couple directly to $\pi$. Hence the
gravitational potential relevant for their motion must be suppressed by a factor of $1/1-g$ relative to the Newtonian potential.

The expression for $\Phi_-$ follows from substituting~(\ref{phi-}) into~(\ref{modpoisson}):
\bea
\nonumber
\Phi_-(k) &=& \frac{1}{1+\left(\frac{kr_c}{a}\right)^{2(\alpha-1)}}\cdot \frac{1-\frac{2g}{\epsilon}\left(\sqrt{1+\epsilon}-1\right)}{1-g} \\
& & \qquad \;\;\;\;\;\;\;\;\;\;\;\;\;\;\;\;\;\;\;\;\;\; \cdot 4\pi G\bar{\rho} \frac{a^2}{k^2}\delta_k\,.
\eea
The last factor of $4\pi G\bar{\rho} a^2 \delta_k/k^2$ is what one would use in standard gravity to translate $\delta$ into a lensing potential. Here, however, 
the two prefactors each act to suppress $\Phi_-$ relative to the density perturbation:

\begin{itemize}

\item On scales much larger than the graviton Compton wavelength, such that $kr_c/a\ll 1$, the first prefactor suppresses $\Phi_-$.

\item On highly non-linear scales, such that $\epsilon \gg 1$, we have 
\be
\frac{1-\frac{2g}{\epsilon}\left(\sqrt{1+\epsilon}-1\right)}{1-g}\approx \frac{1}{1-g}\,.
\ee

\end{itemize}
Since $g\rightarrow -1/3$ at late times, the local Vainshtein effect therefore
results in a 4/3 suppression factor for $\Phi_- $ on these scales.

%%%%%%%%%%%%%%%%%%%%%%%%%%%%%%%%%%%%%%%%
\section{Conclusions}\label{conc}

Over the next decade observations of the large scale structure will subject the $\Lambda$CDM/standard gravity paradigm, thus far
remarkably successful at accounting for cosmological data, to increasingly stringent tests of its predictions. In this work we have studied the possibility
that cosmic acceleration instead stems from a breakdown of Einstein gravity at cosmological distances, due to the graviton having a small mass.

Specifically, we have focused on the normal branch of the DGP model as well as its extension to higher dimensions, called cascading gravity. 
The longitudinal mode of the graviton in these models results in stronger gravitational attraction which, thanks to the Vainshtein screening effect, only kicks in at late times
and on sufficiently large scales. We have focused exclusively on the ``normal" branch of these theories, which are ghost-free
and {\it not} self-accelerating. (Note that the effects uncovered in this work would be exactly the opposite on the self-accelerating branch --- the longitudinal mode would mediate a repulsive force, thereby impeding the formation of structure.)

We have presented a simple phenomenological procedure for calculating the development
of large scale structure in the standard DGP model and its higher-dimensional, cascading gravity cousins.
To focus on the effects of modifications to the growth history, we have assumed a background $\Lambda$CDM expansion history in all cases.
This should be a good approximation for the actual background in cascading models, where the corrections to the Friedmann equation is expected to 
be a slowly-varying function of $Hr_c$.

Our N-body simulations confirm the expectation that structure is more evolved in modified gravity than in $\Lambda$CDM cosmology:
\begin{itemize}
\item Structure grows faster on large scales, leading to enhanced clustering for fixed primordial normalization.
\item As a converse, for a power spectrum normalized to the power observed {\it today}, 
our model predicts systematically less power in the past than
would be inferred from a standard $\Lambda$CDM evolution.
\item Nonlinear processing reaches longer length scales thanks to enhanced gravitational strength,
leading to an additional enhancement in power on nonlinear scales relative to standard gravity.
\item Nonetheless, freedom to marginalize over galaxy bias allows our modified gravity model to fit
the power spectrum derived from the Sloan Digital Sky Survey main galaxies. 
\end{itemize}
While our understanding of these simple results are encouraging, they represent only the
beginning of our program for testing this class of modifications of gravity. Galaxy power
 spectra, with their bias uncertainty, are not the most discriminating measurements for constraining
or demonstrating the particular effects of these models. 

Several observational probes can constrain or confirm the phenomenology of our class of models.
For instance, bulk flows and weak lensing are sensitive directly to collections of dark matter.
Since gravity does not lens light with the same potential as it attracts matter in our 
model, though, lensing will not be as useful in constraining this model as it may appear at first.
Bulk flows, though, give a direct measure of the local matter potential, and are becoming a 
more robust observational tool. Intriguingly, recent results from the compilation of
peculiar velocity surveys \cite{Feldman:2008rm} and systematically uncertain, but
highly suggestive, kinetic Sunyaev-Zel'dovich observations \cite{Kashlinsky:2008ut} 
have provided evidence that the peculiar motions of galaxies on the largest length
scales are higher than that expected in the standard cosmological paradigm.
Our model may be able to account for these enhanced flows \cite{prep}.

\section*{Acknowledgements}
The authors would like to thank Niayesh Afshordi, Neal Dalal, Ghazal Geshnizjani, Wayne Hu, Mike Hudson, Bhuvnesh Jain, Marcos Lima, Pat McDonald, James Peebles, Roman Scoccimarro, Fritz Stabenau, Andrew Tolley and Mark Trodden for enlightening discussions. We are also grateful to the anonymous Referee for encouraging us to spell out the regime of validity of~(\ref{modpoisson}) in Sec.~\ref{continuum}. This work was supported in part by the Perimeter Institute for Theoretical Physics.  Research at Perimeter Institute is supported by the Government of Canada through Industry Canada and by the Province of Ontario through the Ministry of Research \& Innovation.

\appendix
\section{Relation to Decoupling Arguments}
\label{decoup}

In this Appendix we elucidate the relation between our modified Poisson equation and the $\pi$ equation of motion derived in a decoupling
limit of DGP~\cite{strong,luty,nicolis}. For simplicity, we neglect the effect of the cosmological background and treat the source as embedded
in flat Minkowski space. Hence, $g\rightarrow -1/3$ and $\bar{\rho}\,\delta \rightarrow \rho$. Moreover, we focus on scales much smaller than the graviton Compton wavelength: $kr_c \gg 1$. In this approximation,~(\ref{modpoisson}) reduces to
\be
\nabla^2\Psi = 4\pi G\rho \left(1+\frac{9}{32\pi Gr_c^2\rho}\left[\sqrt{1 + \frac{64\pi G r_c^2\rho}{27}}-1\right]\right)\,.
\ee
The contribution from $\pi$ in the above is the difference from the usual Poisson equation, namely
\bea
\nonumber
\nabla^2 \pi  &=& 4\pi G\rho \frac{9}{32\pi Gr_c^2\rho}\left[\sqrt{1 + \frac{64\pi G r_c^2\rho}{27}}-1\right] \\
&=& \frac{9}{8r_c^2} \left[\sqrt{1 + \frac{64\pi G r_c^2\rho}{27}}-1\right]\,.
\label{pi1}
\eea
In particular, in the limit $Gr_c^2\rho \ll 1$, this simplifies to
\be
\nabla^2\pi = \frac{4\pi G}{3}\rho\,.
\ee
The factor of $1/3$ relative to the standard Poisson equation is consistent with the correction in~(\ref{weak}).

Let us compare these results with the $\pi$ equation of motion of DGP. In a certain decoupling limit of the theory~\cite{strong,luty,nicolis}, 
the longitudinal mode decouples and obeys the equation of motion 
\be
3\nabla^2\pi + \frac{2}{r_c^2}\left[(\nabla^2\pi)^2-(\nabla_i\nabla_j\pi)^2\right]= 4\pi G\rho\,.
\label{pi2}
\ee
In general, because of the $\nabla_i\nabla_j\pi$ term, we cannot algebraically solve for $\nabla^2\pi$, as implied
by~(\ref{pi1}).

\begin{figure}[h] %  figure placement: here, top, bottom, or page
   \centering
   \includegraphics[width=0.5\textwidth]{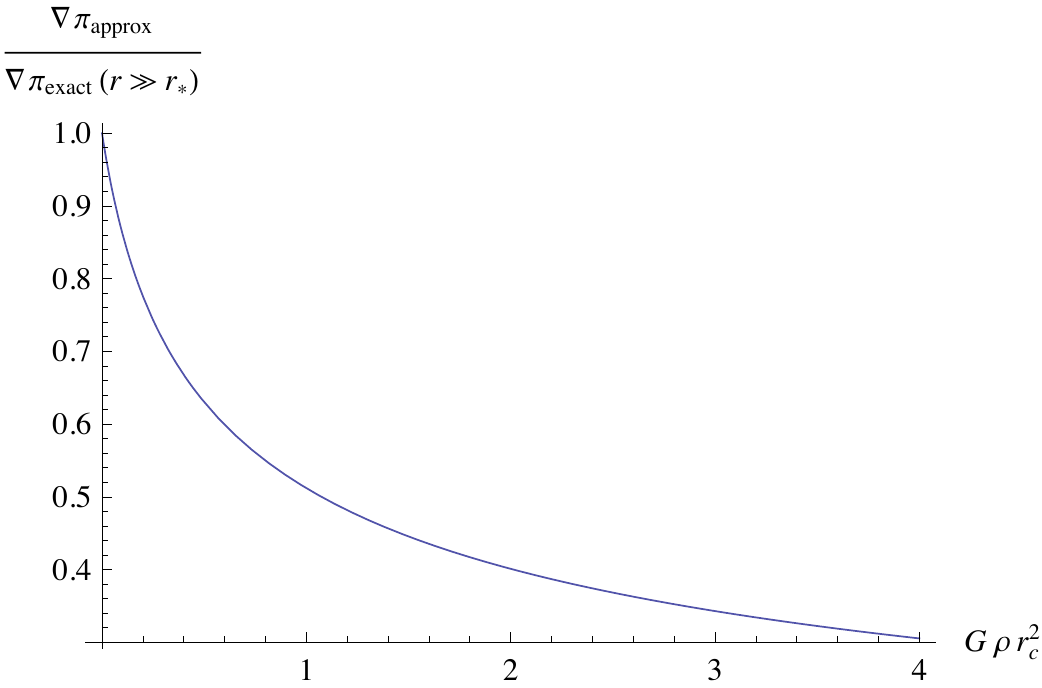} 
   \caption{Comparison of our approximate solution for $\pi$ with exact profile in the far-field regime. We do not expect this mismatch to be 
   very consequential at the level of approximation used in this study because most of the dynamics in our simulation are driven by regions
   whose overdensity $G \rho r_c^2\ll 1$, where the approximate and exact solution are most similar. }
      \label{ratiopi}
\end{figure}

However, within a spherical top-hat of radius $R$, it is easily seen that the solution is given by
\be
\pi = \frac{3r^2}{16r_c^2}\left[\sqrt{1 + \frac{64\pi G r_c^2\rho}{27}}-1\right]\;\;\; {\rm for}\; r\leq R\,,
\ee
which agrees precisely with~(\ref{pi1}). After all this is not surprising since we obtained~(\ref{pi1}) from
the spherical top-hat solutions of Lue {\it et al.}~\cite{lue}. At the origin of this agreement is the fact
that $\pi(r)\sim r^2$ within the top-hat, and hence 
\be
(\nabla_i\nabla_j\pi)^2 = \frac{1}{3}(\nabla^2\pi)^2\,.
\label{key}
\ee
Therefore, the underlying assumption in~(\ref{pi1}) is {\it that~(\ref{key}) holds in general}.

To see what this entails, consider the solution outside the top-hat ($r>R$). From~(\ref{pi1}), we obtain
\be
|\nabla\pi_{\rm approx}| = \frac{GM}{3r^2}\cdot \frac{27}{32\pi G r_c^2\rho}\left[\sqrt{1 + \frac{64\pi G r_c^2\rho}{27}}-1\right]\,.
%\frac{{\rm d}\pi_{\rm approx}}{{\rm d}r} = \frac{GM}{3r^2}\cdot \frac{27}{32\pi G r_c^2\rho}\left[\sqrt{1 + \frac{64\pi G r_c^2\rho}{27}}-1\right]\,.
\ee
The solution to the exact equation~(\ref{pi2}), on the other hand, decreases more slowly, since $|\nabla\pi_{\rm exact}| \sim 1/r^{1/2}$ for
$r < r_\star$, as can be seen from~(\ref{modpot}), and approaches the $1/r^2$ profile asymptotically: $|\nabla\pi_{\rm exact}|
\rightarrow GM/3r^2$ for $ r\gg r_\star$. 

In other words, far away from the source we have
\be
\left\vert \frac{\nabla\pi_{\rm approx}}{\nabla\pi_{\rm exact}}\right\vert_{r\gg r_\star} = \frac{27}{32\pi G r_c^2\rho}\left[\sqrt{1 + \frac{64\pi G r_c^2\rho}{27}}-1\right]\,.
\ee
For small density, $G\rho r_c^2 \ll 1$, this ratio approaches unity. For large density, on the other hand, our approximation tends to underestimate the
far-field effect of $\pi$. This ratio is plotted in Fig.~\ref{ratiopi} as a function of $G\rho r_c^2$. Hence, our simulations offer a conservative approximation
to the actual modifications induced by the scalar-mediated force.

\section{Modifications to Non-Linear Fitting Algorithm}
\label{smithfit}

The changes to the Smith {\it et al.} fitting coefficients necessary to fit our runs are as follows. Any parameters not mentioned are left
as in the original algorithm. Since we have only one set of simulations, we cannot give the full dependence of the parameters
on the effective index, $n$, or the spectral curvature; these dependencies will, for our purposes, be left the same as in 
the original algorithm. The one alteration we make that is not part of the original algorithm is to change the definition of the
parameter $y$ from $y(\mbox{Smith}) \equiv k / k_\sigma$ to $y(\mbox{mod}) \equiv \eth k / k_\sigma$; that is, 
we introduce a new parameter, $\eth$.

The relevant parameters and their modified values are listed in Table~\ref{smithparams}.
These parameters have been calibrated only for our small sample of runs. A code containing
the modified Smith algorithm, with updated fitting coefficients if we perform a wider variety of runs,
will be made available~\cite{code}. We provide a single set of parameters because
both the DGP and Degravitation simulations can both be fit reasonably well by them. There is some
disagreement for the smallest values of $r_c$ (see Fig. \ref{dgpvdeg}), though. To fit these, 
it is necessary to use slightly different parameters for each case. For Degravitation, the small $r_c$ 
fit requires $\eth=1.02$, the correction to $\beta$ to be $2.0$, and the correction to $\mu_n$ and 
$\nu_n$ to be $0.8$. For DGP, the small $r_c$  fit requires $\eth=1.05$, the correction to $\beta$ to be $1.9$,
and the correction to $\mu_n$ and  $\nu_n$ to be $0.95$.

\begin{table}[htb]
\small
\begin{tabular}{|c|c|}
\hline
%\hspace{5pt}                Modified gravity      & \\
%\hline
\hspace{5pt}$\log_{10} a_n$                      & $0.84 \log_{10} a_n^{\rm Std}$ \\
\hspace{5pt}$\log_{10} b_n$                    &  $\log_{10} b_n^{\rm Std} + \log_{10} 1.1$ \\
\hspace{5pt}$\log_{10} c_n$                  & $\log_{10} c_n^{\rm Std} + \log_{10} 1.05$  \\
\hspace{5pt}$\log_{10} \mu_n$                                 & $\log_{10} \mu_n^{\rm Std} + \log_{10} 0.875$ \\
\hspace{5pt}$\log_{10} \nu_n$                                 &   $\log_{10} \nu_n^{\rm Std} + \log_{10} 0.875$\\
\hspace{5pt}$\alpha_n$					    & $0.8 \,\alpha_n^{\rm Std}$ \\
\hspace{5pt}$\beta_n$					    & $1.95 \, \beta_n^{\rm Std}$ \\
\hspace{5pt}$\eth$					    & 1.035\\
\hline
\end{tabular}
\caption{Parameters for the modified non-linear fitting algorithm. The superscript ``Std" indicates the value of the parameter
in the standard Smith {\it et al.} formula. \label{smithparams}}
\end{table}

%
%\begin{tabular}{|c|c|c|}
%\hline
%\hspace{5pt}                             & Degravitation ($\alpha = 0$)  & DGP ($\alpha = 0.5$)  \\
%\hline
%\hspace{5pt}$\log_{10} a_n$                      & $0.84 \log_{10} a_n^{\rm Std}$ &  $0.84 \log_{10} a_n^{\rm Std}$   \\
%\hspace{5pt}$\log_{10} b_n$                    &  $\log_{10} b_n^{\rm Std} + \log_{10} 1.1$  &  $\log_{10} b_n^{\rm Std} + \log_{10} 1.1$ \\
%\hspace{5pt}$\log_{10} c_n$                  & $\log_{10} c_n^{\rm Std} + \log_{10} 1.05$ &  $\log_{10} c_n^{\rm Std} + \log_{10} 1.05$ \\
%\hspace{5pt}$\log_{10} \mu_n$                                 & $\log_{10} \mu_n^{\rm Std} + \log_{10} 0.8$  & $\log_{10} \mu_n^{\rm Std} + \log_{10} 0.95$  \\
%\hspace{5pt}$\log_{10} \nu_n$                                 &   $\log_{10} \mu_n^{\rm Std} + \log_{10} 0.8$ &  $\log_{10} \mu_n^{\rm Std} + \log_{10} 0.95$  \\
%\hspace{5pt}$\alpha_n$					    & $0.8 \,\alpha_n^{\rm Std}$  & $0.8 \,\alpha_n^{\rm Std}$ \\
%\hspace{5pt}$\beta_n$					    & $2.0 \, \beta_n^{\rm Std}$  & $1.9 \, \beta_n^{\rm Std}$  \\
%\hspace{5pt}$\eth$					    & 1.02 & 1.05  \\
%\hline
%\end{tabular}
%\caption{Parameters for the modified non-linear fitting algorithm. The superscript ``Std" indicates the value of the parameter
%in the standard Smith {\it et al.} formula. \label{smithparams}}
%\end{table}

\end{document}